\documentclass[aps,prd,reprint, superscriptaddress,preprintnumbers, longbibliography, amsmath,amssymb,amsfonts, nofootinbib]{revtex4-1}
\usepackage{graphicx}
\usepackage[dvipsnames]{xcolor}
\usepackage{hyperref}
\usepackage{xspace}
\usepackage{ifdraft}
\usepackage{epstopdf}
\usepackage{slashed}

\usepackage{tikz-feynman} 

\tikzfeynmanset{compat=1.1.0}
\tikzfeynmanset{/tikzfeynman/warn luatex = false}

\definecolor{nhpRed}{RGB}{161,0,0}
\definecolor{nhpBlue}{RGB}{0,100,144}
\definecolor{jaxoblue}{HTML}{0086FF}
\definecolor{cutred}{RGB}{219,56,49}
\definecolor{hgreen}{RGB}{25,176,146}
\definecolor{hgreen1}{RGB}{175,230,175}
\definecolor{hblue}{RGB}{52,152,219}
\definecolor{hblue1}{RGB}{255,255,166}
\definecolor{hred}{RGB}{216,83,117}
\definecolor{hred1}{RGB}{255,155,155}
\definecolor{cutred}{RGB}{219,56,49}
\definecolor{hgrey4}{RGB}{75,75,75}
\definecolor{hgrey5}{RGB}{50,50,50}
\definecolor{hgrey3}{RGB}{100,100,100}
\definecolor{hgrey}{RGB}{125,125,125}
\definecolor{hgrey2}{RGB}{125,125,125}
\definecolor{hgrey1}{RGB}{150,150,150}
\definecolor{hgrey0}{RGB}{175,175,175}
\definecolor{darkgreen}{RGB}{255,153,153}

\usepackage{xcolor}
\hypersetup{
    colorlinks,
    linkcolor={nhpRed},
    citecolor={nhpBlue},
    urlcolor={nhpBlue}
}

\definecolor{cutred}{RGB}{219,56,49}

\newcommand{\compSym}[2]{\left(#1\,\textcircled{$\mathfrak{d}$}\,#2\right)}

\definecolor{NUpurple}{RGB}{078,042,132}

\def\sect#1{section~{\ref{#1}}}

\def\spa#1.#2{\left\langle#1\,#2\right\rangle}
\def\spb#1.#2{\left[#1\,#2\right]}
\def\spash#1.#2{\spa{\smash{#1}}.{\smash{#2}}}
\def\spbsh#1.#2{\spb{\smash{#1}}.{\smash{#2}}}
\def\sand#1.#2.#3{%
\left\langle\smash{#1}{\vphantom1}^{-}\right|{#2}%
\left|\smash{#3}{\vphantom1}^{-}\right\rangle}
\def\sandpp#1.#2.#3{%
\left\langle\smash{#1}{\vphantom1}^{+}\right|{#2}%
\left|\smash{#3}{\vphantom1}^{+}\right\rangle}
\def\sandpm#1.#2.#3{%
\left\langle\smash{#1}{\vphantom1}^{+}\right|{#2}%
\left|\smash{#3}{\vphantom1}^{-}\right\rangle}
\def\sandmp#1.#2.#3{%
\left\langle\smash{#1}{\vphantom1}^{-}\right|{#2}%
\left|\smash{#3}{\vphantom1}^{+}\right\rangle}

\def\Tr{\, {\rm Tr}}
\def\tr{\, {\rm tr}}

\def\nn{\nonumber}

\def\eqn#1{eq.~(\ref{#1})}

\def\be{\begin{equation}}
\def\ee{\end{equation}}
\def\bea{\begin{eqnarray}}
\def\eea{\end{eqnarray}}
\def\ba{\begin{eqnarray}}
\def\ea{\end{eqnarray}}

 \definecolor{MattOrange}{rgb}{1.0,0.4,0.2}

\begin{document}

\preprint{
}

\author{John Joseph M. Carrasco}
\affiliation{Department of Physics and Astronomy, Northwestern
  University, Evanston, Illinois 60208, USA}
\affiliation{Institut de Physique Th\'{e}orique, Universite Paris Saclay, CEA, CNRS, F-91191 Gif-sur-Yvette, France}
\author{Nicolas H. Pavao}
\affiliation{Department of Physics and Astronomy, Northwestern
  University, Evanston, Illinois 60208, USA}

\title{Virtues of a symmetric-structure double copy}

\begin{abstract}
 We demonstrate a physical motivation for extending color-dual or BCJ double-copy construction to include theories with kinematic numerators that obey the same algebraic relations as symmetric structure constants, $d^{abc}=\text{Tr}[\{T^{a},T^{b}\}T^c]$. We verify that $U(N_c)$ nonlinear sigma model (NLSM) pions, long known to be color-dual in terms of antisymmetric adjoint factors, $f^{abc}$, are also color-dual in the sense of symmetric color structures, $d^{abc}$, explicitly through six-point scattering. This reframing of NLSM pion amplitudes complements our compositional construction of $d^{abc}$ color-dual higher derivative gauge operators. With adjoint and symmetric color-dual kinematics, we can span all four-point effective photon operators via a double-copy construction using amplitudes from physical theories. We further comment on a tension between locality and adjoint effective numerators, and the implications for spanning gravitational effective operators with non-adjoint kinematics.  
\end{abstract}

\maketitle

\section{Introduction}
Probing for footprints of UV physics with the fields relevant to the available IR scales via effective field theory (EFT) methods is by now a well oiled machine \cite{Georgi:1993mps}. Model building with EFT amounts to enumerating all possible operators consistent with observable IR symmetries, and capturing signatures of the UV in \textit{a priori} independent Wilson coefficients. In recent years EFTs have been constructed to describe phenomena among a wide range of physical scales, from high energy particle physics \cite{Weinberg:1979sa,Buchmuller:1985jz, Isgur:1989vq, Bauer:2002nz,Brivio:2017vri},  to classical gravitational binary inspiral~\cite{Goldberger:2004jt, Porto:2005ac,Cheung:2018wkq, Edison:2022cdu}, to  cosmological inflation \cite{Creminelli:2006xe,Cheung:2007st,Weinberg:2008hq,Burgess:2009ea},  dark energy \cite{Gubitosi:2012hu}, and large scale structure  \cite{Baumann:2010tm,Carrasco:2012cv, Porto:2013qua}.

One of the primary universal challenges in EFT construction is identifying a minimal basis of operators needed at a particular mass-dimension satisfying the desired symmetries. A partial solution has been provided via Hilbert Series methods, see e.g.~refs.~\cite{Lehman:2015via,Henning:2015daa} and references therein, to count the requisite operators.  Identifying the actual operators and their scattering predictions requires more work. For many of the computational problems facing precision calculations in quantum field theory,  on-shell methods have opened new pathways. Indeed, many surprising properties of EFTs have been clarified \cite{Elvang:2018dco,Shadmi:2018xan,Durieux:2019eor,Arkani-Hamed:2020blm,Caron-Huot:2021rmr,Bern:2021ppb, Bern:2022yes} by computing directly at the level of scattering amplitudes, circumventing conventional Feynman rule techniques. 

The past few decades have seen on-shell methods applied to a number of quantum field theories, from the formal to the phenomenological, leading the modern amplitudes program to identify perturbative structure of scattering  that have lead to remarkable reductions in the computational complexity of the S-matrix. One such example, first identified in tree-level gluon amplitudes~\cite{BCJ}  and later generalized to the multi-loop level~\cite{BCJLoop}, is the duality between adjoint color and kinematics, and the associated double-copy construction of amplitudes in a wide range of field theories. Double-copy construction has notably been used  to constrain the Wilson coefficients of higher-derivative operators in gravity theories~\cite{Bern:2017puu,Bern:2017tuc,Bern:2021ppb}.   Indeed the adjoint duality between color and kinematics has been generalized~\cite{Carrasco:2019yyn,Low:2019wuv,Low:2020ubn,Carrasco:2021ptp} to admit non-adjoint color-factors that combine with kinematics to build adjoint-type building blocks in double copy construction that obviates the need for ansatze and can climb all the way to the UV via composition.  

Certain desired adjoint-type EFT building blocks, however, can be confusing when used at low mass-dimension --- suggesting unphysical massless higher-spin exchange~\cite{Pavao:2022kog}. We will show that generalizing the duality between color and kinematics to include purely symmetric color weights, beyond traditional adjoint-color, can resolve challenges to physically consistent double-copy construction. Here we demonstrate the possibilities by considering photonic $U(1)$ duality preserving higher derivative operators.

To realize the traditional adjoint correspondence between color and kinematics, one expresses an $n$-point gauge theory amplitude as a sum over trivalent (cubic) graphs, 
\begin{equation}
    \mathcal{A}_n= \sum_{g\in \Gamma^{(3)}_n} \frac{c_g n_g}{d_g}\, ,
\end{equation}
where for any graph in the set, $\Gamma^{(3)}_n$, of unordered $n$-point cubic graphs, $c_g$ are the color weights, $n_g$ are the kinematic part of the numerator, and $d_g$ are the propagators that encode the local structure of the amplitude. In this form, a gauge theory is said to be \textit{color-dual} if a gauge can be chosen such that the kinematic numerators, $n_g$, obey the same algebraic constraints as the color factors, $c_g$. As an explicit example, consider a four-point color factor functionally defined as:
\begin{equation}
    c(1,2,3,4) = f^{a_1a_2 e}f^{e a_3 a_4}
\end{equation}
where the structure constants can be written in terms of $U(N_c)$ group theory generators as $f^{abc} \equiv \text{Tr}[[T^{a},T^{b}]T^c]$. Such color weights have a variety of symmetry properties that it inherits from the underlying color algebra, namely \textit{anti-symmetry} and satisfying the \textit{Jacobi identity}, 
\begin{align}
\label{colorRel}
    c(1,2,3,4)&+ c(1,2,4,3)=0
    \\
    c(1,2,3,4)&+c(1,3,4,2)+c(1,4,2,3)=0
\end{align}
A theory is said to be \textit{color-dual} when the kinematic numerator for any graph shares the same algebraic properties as the corresponding color factor. For the four-point color factor above, that amounts to finding a set of kinematic numerators such that,
\begin{align}
    n(1,2,3,4)&+ n(1,2,4,3)=0
    \\
    n(1,2,3,4)&+n(1,3,4,2)+n(1,4,2,3)=0
\end{align}
Many such theories satisfy this property; for recent reviews of the topic, see \cite{BCJReview,Bern:2022wqg, Adamo:2022dcm}. An added feature of this construction is that since gauge invariance in the amplitude, $\mathcal{A}_n$, is encoded in the algebraic relations between the color factors, replacing the color weights, $c_g$, with color-dual numerators, $\tilde{n}_g$, must still yield gauge invariant amplitudes, albeit for a different theory, of the form:
\begin{equation}
\label{dcAmp}
    \mathcal{M}_n = \sum_{g\in \Gamma^{(3)}_n} \frac{\tilde{n}_g n_g}{d_g}
\end{equation}
This procedure is known as double copy construction~\cite{Bern:2010ue}, and theories whose tree-level amplitudes can be described in such a form are said to be double-copy constructible. While the first realization of this construction arose from the study of multi-loop amplitudes from $\mathcal{N}=8$ supergravity, many theories, including the effective field theory of Born-Infeld (BI) photons, have been shown to permit such a construction \cite{Cachazo2014xea}.  

With this in hand, the problem of double-copy construction of local QFT observables in a wide web of theories is reduced to identifying color-dual functions of on-shell kinematic variables. Through a simple adjoint composition rule, previous studies have identified all higher-derivative scalar and vector (single-trace) numerators that obey adjoint-type kinematic relations at four-points~\cite{Carrasco:2019yyn} as well as all such scalar building blocks at five-points~\cite{Carrasco:2021ptp}. Using the double-copy, these  color-dual building blocks can efficiently encode a vast landscape of gauge theory and gravity operators, many of which may be needed for good UV behavior. 

As we discuss in this work, however, the most physically meaningful double-copy construction may not always involve the adjoint.  For example, consider the four-field photon operator of the form:
\begin{equation}
\label{BIcT}
    \widehat{\mathcal{O}}^{(++++)} = \int d^4 x \left[(\partial_\mu F^{+}_{\dot{\alpha}\alpha})(\partial^\mu F_{\alpha\dot{\alpha}}^+)\right]^2
\end{equation}
where $\alpha$, $\dot{\alpha}$ are spinor indices of the positive-helicity chiral field strength, defined as,
\begin{equation}
    F^+=\frac{1}{2}(F^{\mu\nu}+i\tilde{F}^{\mu\nu})\sigma_{\mu\nu}
\end{equation}
where $\sigma_{\mu\nu} = \frac{1}{2}(\sigma_\mu\Bar{\sigma}_\nu-\sigma_\nu\Bar{\sigma}_\mu)$, and $\tilde{F}_{\mu\nu}=\epsilon_{\mu\nu\rho\sigma}F^{\rho\sigma}$ is the dual field strength. In \cite{Elvang:2020kuj} it was shown that this operator is required to cancel the $U(1)$ anomaly in the Born-Infeld S-matrix through one-loop. While BI theory is itself double-copy constructible from Yang-Mills and NLSM, this counterterm does not appear to be constructible from local four-field higher-derivative operators added to either Yang-Mills or the NLSM.   This operator corresponds to a local four-point photon amplitude,
\be
\label{symOP}
    \mathcal{O}^{(++++)} = \sigma_2^2\,\mathcal{T}
\ee
where we introduce the standard permutation invariant all-plus tensor structure:
\begin{equation}
    \mathcal{T} = \frac{[12][34]}{\langle 12\rangle\langle 34\rangle}\,
\end{equation}
and the scalar permutation invariant $\sigma_2 = s^2+ t^2+u^2$.  The authors of ref.~\cite{Elvang:2020kuj} demonstrated that there did not exist local higher-derivate ordered amplitudes that satisfied the $(n-3)!$ BCJ relations  which could double-copy to this amplitude.  

This paper is organized as follows. As we will explain in \sect{nonAdjNonLocal}, it is possible to build this operator via adjoint double-copy using the vector building blocks of \cite{Carrasco:2019yyn, Carrasco:2021ptp}, through a somewhat surprising loophole.  The required higher-derivative gauge theory's four-point amplitude sits on non-trivial (presumably unphysical) higher-spin  particle exchange.  These channels cancel in the double-copy construction to land on the local prediction for the higher-derivative photon exchange.  

If all we cared about was constructing a particular counterterm via the double-copy, then this might be fine; but our priority in looking at this is to develop an understanding of when and how double-copy EFT can be built from physical theories.  

This paper is organized as follows. We will show in \sect{symmDoubleCopy} that generalizing double-copy construction to admit $d^{abc}$ type color-kinematics duality is key to the  generation of operators of the type \eqn{symOP} from physically consistent theories.  In \sect{nonAdjNonLocal} we provide details about the unphysical character of particular adjoint representations. We further show in \sect{sect:four} that a small combination of adjoint and symmetric operators encode the $U(1)$ anomalous tensor structures through two-loops in duality invariant electromagnetism. In \sect{discussion}, we conclude by discussing the opportunities for future work and generalizations.

%
%
\section{Local counterterms with symmetric color-structures}
\label{symmDoubleCopy}
It is worthwhile to review explicitly how Born-Infeld photon amplitudes can be generated using the adjoint double copy. Since photons are colorless vectors, we anticipate that their amplitudes can be written as in \eqn{dcAmp}.  By little group scaling, if one of the kinematic numerators belongs to a vector theory, the other kinematic numerator will belong to a scalar theory.  The mass-dimension of Born-Infeld forces the kinematic numerators have to be order-two in mandelstam variables.  At four-points, requiring color-kinematics duality uniquely restricts us to  NLSM pions for the scalar theory \cite{Carrasco:2019qwr}. Indeed the Adler-zero property of NLSM translates to the self-duality of Born-Infeld amplitudes through double-copy. Explicitly, Born-Infeld photon amplitudes, $ \mathcal{A}^{\text{BI}}$, can be constructed from adjoint double copies as follows:
\begin{align}
\label{eqs:BI4ptManip}
    \mathcal{A}^{\text{BI}}_4 &= \mathcal{A}_4^{\text{YM}} \otimes \mathcal{A}_4^{\text{NLSM}} \\
                      &\equiv \sum_{g\in \Gamma^{(3)}} \frac{ n^{\text{YM}}_g  n^{\pi}_g}{d_g} \nonumber \\
                      &\propto \frac{ ( s t A^\text{YM}_{(s,t)} ) ( s t A^\text{NLSM}_{(s,t)})}{s t u} \nonumber\\
                      &= s t A^\text{YM}_{(s,t)} 
\end{align}
where $\otimes$ denotes the generalized product of kinematic numerators between the two amplitudes in the double copy.  The last two lines of \eqn{eqs:BI4ptManip} exploit Jacobi and antisymmetry of numerators to write the four-point double-copy amplitude in terms of the $s,t$ channel ordered amplitudes, $A_{(s,t)}$, of Yang-Mills and NLSM, and the fact that $A_{(s,t)}^\text{NLSM} \propto u$. 
If we want to describe higher-derivative adjoint double-copy operators, we can trivially consider powers of scalar permutation invariants associated with either numerator (which will disrupt neither Jacobi relations nor gauge invariance), or consider higher derivative scalar building blocks, or allow vectors to have higher-derivative weights.  See e.g. refs.~\cite{Carrasco:2019yyn,Carrasco:2021ptp} for constructions along these lines.  

 At four-points there are only eight distinct vector building blocks (up to powers of scalar permutation invariants), and only two distinct scalar building blocks, so the ansatz required to span any given adjoint-constructible four-point higher-derivative operator at any mass-dimension over BI is relatively straightforward.  Note that a consequence of the manipulations demonstrated in \eqn{eqs:BI4ptManip} is that should \eqn{symOP} be double-copy constructible, then the required higher-derivative gauge-amplitude is simply:
 \begin{align}
 \label{eqn:necAdjSingleCopy}
  A^{\text{HD}}_{(s,t)} &= \mathcal{T} \frac{\sigma_2^2}{s \, t} \\
    &= 2\, \mathcal{T}  \left(\frac{t^3}{s} + \frac{s^3}{t}  + \sigma_2 + s\, t \right) \nn.
 \end{align}
 This ordered amplitude has higher-spin residues on both $s$ and $t$ channel poles~\cite{Pavao:2022kog}, so it cannot correspond to a local four-point operator.  Amusingly it does satisfy the BCJ relations and is constructible from  the vector building blocks of~\cite{Carrasco:2019yyn,Carrasco:2021ptp}, but these vector building blocks are typically be used with at least one power $\sigma_3\equiv s t u$ to describe local gauge-theory counterterms.

We appear to have hit a major road block threatening to derail the double-copy construction of operators like \eqn{symOP} from local theories. However, NLSM has a novel color-dual property that is obscured by the typical approach to writing its color-weights in terms of adjoint color factors as we will now describe.

\paragraph*{New Color Basis for NLSM:} 

First, consider the NLSM Lagrangian up to leading order:
\begin{multline}
    \mathcal{L}^{\text{NLSM}} = \frac{1}{2}(\partial \phi)^a(\partial \phi)^a+\frac{\Lambda}{2} f^{abe}f^{ecd}(\partial\phi)^a\phi^b\phi^c (\partial\phi)^d
    \\
    + \mathcal{O}(\Lambda^2)
\end{multline}
Using the Fierz Identity for $SU(N_c)$ group theory generators,
\begin{equation}
    (T^a)_{ij}(T^a)_{kl} = \delta_{jk}\delta_{il}-\frac{1}{N_c}\delta_{ij}\delta_{kl}
\end{equation}
the structure constants above can be re-expressed in terms of color traces:
\begin{multline}
    f^{abe}f^{ecd} = \Tr[T^a T^b T^c T^d]-\Tr[T^a T^b T^d T^c]
    \\+\Tr[T^a T^d T^c T^b]-\Tr[T^a T^c T^d T^b]
\end{multline}
At four points we can define the following color-factors for the three cubic graphs:
\begin{equation}
\begin{split}
    c_s^{\text{ff}} &\equiv f^{a_1a_2e}f^{ea_3a_4} 
    \\
    c_t^{\text{ff}} &\equiv f^{a_1a_4e}f^{ea_3a_2} 
    \\
    c_u^{\text{ff}} &\equiv f^{a_1a_3e}f^{ea_2a_4} 
   \end{split}
\end{equation}
The definition of color factors in terms of linearly independent color traces can make manifest the algebraic relations specified in \eqn{colorRel}. Exploiting the color Jacobi relation $c^{\text{ff}}_t=c^{\text{ff}}_s-c^{\text{ff}}_u$ we can rewrite the color-dressed scattering amplitude in terms of color-ordered amplitudes:
\begin{align}
\label{eqn:4ptNLSMAmp}
 \mathcal{A}^{\text{NLSM}}_4 &=3\Lambda \left( c^{\text{ff}}_s \,u + c^{\text{ff}}_u \, s\right) \\
  &= \Lambda \left( \frac{c^{\text{ff}}_s n^{\pi,\text{ff}}_s}{s} + \frac{c^{\text{ff}}_t n^{\pi,\text{ff}}_t}{t} + \frac{c^{\text{ff}}_u n^{\pi,\text{ff}}_u}{u} \right) \\
  &= \Lambda \left(  c^{\text{ff}}_s \left( \frac{n^{\pi,\text{ff}}_s}{s} + \frac{n^{\pi,\text{ff}}_t}{t} \right ) + c^{\text{ff}}_u \left( \frac{n^{\pi,\text{ff}}_u}{u} - \frac{n^{\pi,\text{ff}}_t}{t} \right) \right) \\
  &= \Lambda \left(  c^{\text{ff}}_s A^{\text{NLSM}}_{(s,t)}+ c^{\text{ff}}_u A^{\text{NLSM}}_{(t,u)} \right) 
\end{align}
where we can dress cubic graphs with adjoint color-dual numerators
\begin{equation}
\label{eqn:Pion4ptFFNum}
n^{\pi,\text{ff}}_s =s(u-t)=t^2 - u^2
\end{equation}
 and the other channels described by functional relabeling: $n^{\pi, \text{ff}}_t=n^{\pi, \text{ff}}_s|_{s \leftrightarrow t}$ and  $n^{\pi, \text{ff}}_u=n^{\pi, \text{ff}}_s|_{s \leftrightarrow u}$. Of course we could also define \textit{symmetric} color structures in terms of the color generators by introducing an anticommutator as follows:
\begin{equation}
    d^{abc}\equiv \text{Tr}[\{T^a,T^b\}T^c]
\end{equation}
After imposing the Fierz Identity on contractions of $d^{abc}$, one can rewrite the adjoint color factors completely in terms of symmeterized trace structures:
\begin{equation}
\label{adjointToSym}
    f^{ade}f^{ecb} = d^{abe}d^{ecd}-d^{ace}d^{ebd} + \mathcal{O}(1/N_c)
\end{equation}
where the $\mathcal{O}(1/N_c)$ terms are multi-trace delta functions. For simplicity, we will drop these $SU(N_c)$ group theory factor for the remainder of the paper, and restrict ourselves to $U(N_c)$ gauge theories. In this new color basis, after imposing the equations of motion and dropping boundary terms, the NLSM Lagrangian can be re-expressed in the following form
\begin{multline}
    \mathcal{L}^{\text{NLSM}} = \frac{1}{2}(\partial \phi)^a(\partial \phi)^a+\frac{3 \Lambda}{4} d^{abe}d^{ecd}(\partial\phi)^a(\partial\phi)^b\phi^c\phi^d
    \\
    + \mathcal{O}(\Lambda^2)
\end{multline}
The Feynman rules for this Lagrangian are straightforward and yield the \textit{equivalent} color-dressed $U(N_c)$ NLSM amplitude:
\begin{equation}
\label{nlsmSymBasis}
    \mathcal{A}^\text{NLSM}_4 = -3\Lambda \left(c_s^{\text{dd}} s + c_t^{\text{dd}} t+c_u^{\text{dd}} u\right)
\end{equation}
where we have defined the four-point symmetric color factors as:
\begin{equation}
\begin{split}
    c_s^{\text{dd}} &= d^{a_1a_2e}d^{ea_3a_4}
    \\
    c_t^{\text{dd}} &= d^{a_1a_4e}d^{ea_3a_2}
    \\
    c_u^{\text{dd}} &= d^{a_1a_3e}d^{ea_2a_4}
\end{split}
\end{equation}
and the conventions for our Mandelstam invariants are, $s=s_{12}$, $t=s_{23}$, and $u=-(s+t)$. We can see that the above formulation of the NLSM amplitude in a symmetric color basis is equivalent to the perhaps more familiar form of \eqn{eqn:4ptNLSMAmp} by exploiting \eqn{adjointToSym}, $c^{\text{ff}}_t=c^{\text{dd}}_s - c^{\text{dd}}_u$, and relabelings.

Both are indeed permutation invariant by construction; but more importantly, both can be written as a color-dual sum over manifestly local cubic graphs:
\begin{equation}
    \mathcal{A}^{\text{NLSM}} = \sum_{g\in \Gamma^{(3)}} \frac{c^{\text{dd}}_g n^{\pi,\text{dd}}_g}{d_g}=\sum_{g\in \Gamma^{(3)}} \frac{c^{\text{ff}}_g n^{\pi,\text{ff}}_g}{d_g}
\end{equation}
where the color-dual kinematic numerators in the $dd$ color-basis is simply:
\begin{equation}
\label{NLSMnums}
    n^{\pi,\text{dd}}_s = s^2 
\end{equation}
Notice that while the adjoint kinematic-numerator depends on how one chooses to absorb the contact term, the symmetric numerator, $ n^{\pi,\text{dd}}_s$, is invariant under generalized gauge transformations \cite{BCJ}. This redundancy in the adjoint numerators is a measure of algebraic relations between the color factors. At four-point, the symmetric color factors are linearly independent -- a property that does not persist at higher multiplicity. 

We have verified that this feature of NLSM is manifest through six-point. That is, color-dressed NLSM amplitudes can be expressed as a sum over all trivalent graph topologies, weighted by symmetric color factors and kinematic numerators that satisfy the same algebraic identities. This is an important non-trivial check, since the color factors, and also the color-dual kinematic numerators, satisfy additional group theory identities above four-point. 

\paragraph*{Symmetric Vector Numerators:} Given the structure of \eqn{nlsmSymBasis}, we find a path towards double-copy photon operators like in \eqn{symOP} that are invisible to the  double-copy of local  adjoint gauge theory counterterms. To do so, one needs Yang-Mills operators that generate symmetric vector numerators, i.e. kinematics that are color-dual to \eqn{nlsmSymBasis}. One operator that manifests this structure through four-point is:
\begin{align}
\label{LagrangianSymVec}
     \mathcal{L}^{\text{int}} &=  c_{(0,2)}\text{Tr}\left[\{F_{\mu\nu},F^{\mu\nu}\}\{F_{\rho\sigma},F^{\rho\sigma}\}\right]
\end{align}
where $F_{\mu\nu} = F_{\mu\nu}^a T^a$ are $U(N_c)$ field strengths. In four-dimensions, the four-point all-plus vector amplitude generated by this Lagrangian is simply,
\begin{align}
\label{vecSymBasis}
    \mathcal{A}^{\text{dd},2}_{(++++)} &= c_{(0,2)}\mathcal{T}\,(c_s^{\text{dd}} s^2 +c_t^{\text{dd}} t^2+c_u^{\text{dd}} u^2).
\end{align}
First we recognize this can be written in a symmetric color-dual form as:
\begin{align}
    \mathcal{A}^{\text{dd},2}_{(++++)} &= c_{(0,2)} \sum_{g \in \Gamma_4^{(3)}} \frac{c_g^{\text{dd}} n_g^{\text{dd}}}{d_g} 
\end{align}
with 
\be
\label{vecSymNum}
n_s^{\text{dd}} = \mathcal{T} s^3
\ee
and the other channel numerators following the standard relabeling. In \sect{sect:four} we will show that this particular color-dual numerator is a four-dimensional projection of a more general $D$-dimensional spanning set of symmetric vector numerators. 

One should note that, in similar spirit to the scalar kinematics for NLSM, these individual symmetric numerators are independently gauge-invariant. This is in contrast to typical adjoint-vector numerators at four-points.  The adjoint redundancy is a result of algebraic relations between adjoint color-weights where the four-point $c^{\text{dd}}_g$ color weights are independent.  Replacing the color factors with the NLSM symmetric numerators of \eqn{NLSMnums}, yields a symmetric double-copy construction of precisely the matrix element \eqn{symOP} generated by the counterterm in \eqn{BIcT}.

For higher-orders in mass dimension, one can employ a constructive composition rule similar to the adjoint higher-derivative color-dual numerators described in \cite{Carrasco:2019yyn,Carrasco:2021ptp}.  Given two symmetric numerators, $j^{\text{dd}}_g$ and $k^{\text{dd}}_g$, the product maintains their symmetry properties and thus generates a new symmetric numerator:
\begin{equation}
\label{compRule}
    n^{\text{dd}}_s= \compSym{j^{\text{dd}}}{k^{\text{dd}}}_s= j^{\text{dd}}_s k^{\text{dd}}_s\,.
\end{equation}
For scalar kinematics there are linear and quadratic building blocks,
\begin{equation}\label{symmetricBuildingBlocks}
    n^{\text{dd},1}_s = s \qquad n^{\text{dd},2}_s = t\,u
\end{equation}
which can be repeatedly composed with the vector $n^{\text{dd}}_g$ of \eqn{vecSymNum} to achieve arbitrarily high mass-dimension symmetric numerators. Indeed we recover the symmetric color-dual pion numerator, $n_s^{\pi,\text{dd}}$, by composing two factors of the linear building block,
\begin{equation}
n_s^{\pi,\text{dd}}\equiv \compSym{n^{\text{dd},1}}{n^{\text{dd},1}}_s = s^2\,.
\end{equation}

\section{Details on a non-local adjoint double-copy}
\label{nonAdjNonLocal}
Considering the algebraic color relation in \eqn{adjointToSym}, all of the symmetric kinematic numerators above could be expressed in terms of adjoint-type numerators. Using the simple casting rule,
\begin{equation}
    n^{\text{ff}}_s = n^{\text{dd}}_t-n^{\text{dd}}_u\,
\end{equation}
we find candidate adjoint-type numerators:
\begin{equation}
\label{nonLocaladjoint}
    n^{\text{vec,ff}}_s \propto t[14]^2[23]^2-u[13]^2[24]^2\,.
\end{equation}
Note that since,
\begin{align}
\mathcal{T}&=\frac{[12][34]}{\langle12\rangle \langle34\rangle} =\frac{([12][34])^2}{s^2},\\
 &=\frac{([14][23])^2}{t^2} =\frac{([13][24])^2}{u^2}
\end{align}
the adjoint numerator can be expressed as follows,
\be
    n^{\text{vec,ff}}_s=\mathcal{T}(t^3-u^3).
\ee
This is a perfectly fine adjoint color-dual numerator, manifestly antisymmetric around $u\leftrightarrow t$, as well as manifestly consistent with $n_s = n_t+n_u$. Let us consider the ordered amplitude, $A_{(s,t)}$, generated by these numerators:
\begin{align}
    A_{(s,t)}^{\text{vec,ff}} &= \frac{n^\text{vec,ff}_s}{s}+ \frac{ n^\text{vec,ff}_t}{t}\,,\\
      &= 2 \mathcal{T}\left( \frac{t^3}{s} + \frac{s^3}{t} + \sigma_2 + s t \right)\,,\\
      &= \mathcal{T}  \frac{\sigma_2^2}{s t}\,.
\end{align}
We see that we have recovered exactly the ordered amplitude we discovered in \eqn{eqn:necAdjSingleCopy}.  This is the the type of ordered amplitude required to build our counterterm amplitude of \eqn{symOP} via the double-copy with NLSM in the adjoint sense.

Note this is a fine amplitude in every sense other then having a factorization channel at such a high mass-dimension.  This is not in an adjoint sense a local four-field counterterm.  Indeed, considering the argument of ref.~\cite{Pavao:2022kog}, the residue of the $s$-channel pole suggests that there is a spin-3 mode crossing the cut. However, it is happily a local counterterm when expressed in terms of symmetric $d^{abc}$ color-factors with its own double-copy with NLSM to the desired photonic counterterm.

When composing numerators for general theories, locality is important. But for double-copies with NLSM, which has a infinite sequence of pole cancelling contact terms, badly non-local behavior can be hidden by the pion contacts. This example reveals an interesting interplay between adjoint kinematics and locality. Since adjoint kinematic factors possess greater redundancy than their symmetric counterparts, due to satisfying additional jacobi relations, reducing a symmetric color basis to an adjoint one can evidently lead to spurious poles. The available linear relations between amplitudes sets the efficiency of color-dual compression; the same goes for symmetric kinematic building blocks.

\section{Spanning photonic effective counterterms}\label{sect:four}
To emphasize the utility of these symmetric double-copy constructible operators, consider resolving $U(1)$ anomalies at consecutive loop orders in Born-Infeld theory. The classical effective Lagrangian for this theory is,
\begin{equation}
  \alpha'^2\mathcal{L}^{\text{BI}} = 1-\sqrt{1+\frac{(\alpha')^2}{2} \left(F_{\mu\nu}F^{\mu\nu}\right)-\frac{(\alpha')^4}{16} \left(F_{\mu\nu}\tilde{F}^{\mu\nu}\right)^2}\,.
\end{equation}
Born-Infeld theory belongs to a class of duality invariant theories, classified by a tree-level $U(1)$ conservation among the on-shell helicity states. Theories with this symmetry give rise to anomalous matrix elements at one-loop order that do not conserve the $U(1)$ chiral charge \cite{Elvang:2020kuj}. A fascinating application of the above double-copy construction would be to apply the above class of operators to resolve the duality anomaly.  Indeed it is straightforward to line up the loop-level results of \cite{Elvang:2020kuj} with the operators presented above.  Novel results begin at two-loops -- a calculation that is now within reach.

Using the above operator prediction constructions, we can span all four-point higher-derivative photon operators with scalar permutation invariants, a single adjoint vector building block, and two symmetric-type building blocks. 

The requisite adjoint vector building block is labeled as $n^{\text{vec},F^3}$ in \cite{Carrasco:2019yyn}. These generate the gauge-theory amplitude with a single-insertion of an $F^3$ three-point vertex.   All other effective photon operators can be constructed via the symmetric double copy we describe in \sect{symmDoubleCopy}. For this construction, the two symmetric vector building blocks are  of the form,
\begin{equation}
n_s^{\text{vec},\text{dd},1} = f_{12}f_{34} \qquad n_s^{\text{vec},\text{dd},2} = f_{1324}
\end{equation}
where $f_{ijkl}= \text{tr}[F_iF_jF_kF_l]$ and $f_{ij}= \frac{1}{2}\text{tr}[F_iF_j]$ are gauge invariant objects constructed from linearized field strengths, $F_i^{\mu\nu} = k^\mu_i \epsilon^\nu_i - k^\nu_i \epsilon^\mu_i$, that respect the symmetry of the color-dual vector numerators, $n_s^{\text{vec},\text{dd}}$. Since the constructive composition rule does not spoil the gauge invariance of four-point symmetric numerators,  \eqn{compRule} can be used to span the remaining higher-derivative vector building blocks to all orders in mass dimension. 

This gives us an extremely general higher-derivative color-dual amplitude:
\begin{multline}
\label{HDVecAmp}
\mathcal{A}^{\text{vec}+\text{HD}} _4= \sum_{g\in \Gamma^{(3)}}\sum_{x,y} \Bigg[a^{F^3}_{(x,y)}\sigma_3^x \sigma_2^y \frac{n^{\text{vec},F^3}_g c^{\text{ff}}_g }{d_g}
\\
+a^{F^2F^2}_{(x,y)} ({n^{\text{dd},1}_g})^x({n^{\text{dd},2}_g})^y\frac{n^{\text{vec},\text{dd},1}_g c^{\text{dd}}_g  }{d_g}
\\
+a^{F^4}_{(x,y)} ( {n^{\text{dd},1}_g})^x({n^{\text{dd},2}_g})^y \frac{n^{\text{vec},\text{dd},2}_g c^{\text{dd}}_g  }{d_g}\Bigg]
\end{multline}
By performing a numerator level double copy in both the adjoint and symmetric color factors, the most general four-point photon amplitude (essentially that generated from the Euler-Heisenberg Lagrangian) can be stated concisely as,
\begin{equation}\label{HDPhotonAmp}
 \mathcal{M}_4^{\text{photon}}=\mathcal{A}_4^{\text{vec}+\text{HD}}\Big|^{c^{\text{ff}}_g\rightarrow n^{\pi, \text{ff}}_g}_{c^{\text{dd}}_g\rightarrow n^{\text{dd},1}_g}\,,
\end{equation} 
where $n^{\text{dd},1}_g$ is the linear scalar building block of \eqn{symmetricBuildingBlocks}, and thus has the affect of adding a propagator factor to the numerator since $n^{\text{dd},1}_g\equiv d_g$. We have confirmed that this reproduces without redundancy all possible four-point photon operators through $\mathcal{O}(k^{50})$ in mass dimension.

Since the $U(1)$ anomaly is a four-dimensional on-shell symmetry, first it is necessary to see how this spanning set of photon effective operators projects down to four-dimensional helicity states. In the symmetric double copy sector, the non-vanishing helicity configurations contributing to $n_s^{\text{vec},\text{dd},1}$ are,\begin{align}
\tr[F_i^+F_j^+] &=- [ij]^2
\\
\tr[F_i^-F_j^-] &= -\langle ij\rangle^2
\end{align}
and similarly those for $n_s^{\text{vec},\text{dd},2}$ are simply,
\begin{align}
\tr[F_i^+F_j^+F_k^+F_l^+] &=  -\frac{1}{2} s_{ij}s_{jk} \mathcal{T}
\\
\tr[F_i^-F_j^-F_k^+F_l^+] &=  \frac{1}{4} \langle ij\rangle^2[kl]^2
\\
\tr[F_i^-F_j^+F_k^-F_l^+] &=  \frac{1}{4} \langle ik\rangle^2[jl]^2
\\
\tr[F_i^-F_j^-F_k^-F_l^-] &=  -\frac{1}{2} \frac{s_{ij}s_{jk}}{\mathcal{T}}
\end{align}
All other non-vanishing configuration can be constructed via cyclicity of the trace. To get a sense for how known four-dimensional tensor structures can be recovered from our symmetric vector building blocks, we provide the example of the $t_8F^4$ operator, written succinctly with our normalization as,
\begin{equation}
\label{SUSYt8F4}
\frac{1}{2}t_8F^4 = f_{1234}-f_{12}f_{34}+\text{cyc}(2,3,4)
\end{equation}
For completeness, we also provide the four-dimensional non-vanishing helicity amplitudes of $\text{YM}+F^3$ studied in \cite{Broedel2012rc, Bern:2017tuc} that can be constructed exclusively from an adjoint double-copy:
\begin{align}
s t A^{F^3}(1^+2^+3^+4^+)&= - 2 s t u \,\mathcal{T}
\\
s t A^{F^3}(1^+2^+3^+4^-)&= \frac{[12]^2[23]^2[31]^2}{\mathcal{T}}
\end{align}
Now we are prepared to describe how the symmetric kinematic factors can account for the anomalous matrix elements at loop-level. By little group scaling and mass-dimension alone, we know the algebraic part of the one-loop integral must be of the form,
\begin{align}
    \mathcal{A}^{(+++-)}_{\text{1-loop}} &= \mathcal{O}(\epsilon)
    \\
\mathcal{A}^{(++++)}_{\text{1-loop}} &= \mathcal{T}  c^+_{(0,2)}\sigma_2^2+\mathcal{O}(\epsilon)
\end{align}
and likewise the two-loop algebraic part can be expressed as,
  \begin{align}\label{pppmCT2loop}
    \mathcal{A}^{(+++-)}_{\text{2-loop}} &= \frac{([12][23][31])^2}{\mathcal{T}}c^-_{(1,0)} \sigma_3+\mathcal{O}(\epsilon)
    \\    \label{ppppCT2loop}
    \mathcal{A}^{(++++)}_{\text{2-loop}} &= \mathcal{T} \left(c^+_{(2,0)}\sigma_3^2+ c^+_{(0,3)}\sigma_2^3\right)+\mathcal{O}(\epsilon)
\end{align}
Matching the Wilson coefficients in the above expression to our $D$-dimensional expansion in \eqn{HDPhotonAmp}, we find
\begin{equation}
 \label{finalCs}
\begin{split}
c^+_{(0,2)} &=  \frac{1}{8}\left(a^{F^2F^2}_{(2,0)}-a^{F^4}_{(0,1)}\right) \\
c^+_{(0,3)} &= \frac{1}{16}\left(a^{F^2F^2}_{(4,0)}+a^{F^4}_{(0,2)}\right) \\
c^+_{(2,0)} &= \frac{1}{4}\Big(3a^{F^2F^2}_{(4,0)}+3a^{F^2F^2}_{(2,1)}+3a^{F^2F^2}_{(0,2)} \\
&\quad -6a^{F^4}_{(4,0)}-6a^{F^4}_{(2,1)}-6a^{F^4}_{(0,2)}-8a^{F^3}_{(1,0)}\Big) \\ 
c^-_{(1,0)} &= a^{F^3}_{(1,0)}\,.
\end{split}
\end{equation}
We have learned on very general grounds that there is more than enough freedom to cancel the anomalous contributions through two-loops via double-copy constructed operators.  Of course in future studies of the mechanics of quantum duality conservation it will be important  to determine the actual values of $c_{(x,y)}$ for Born-Infeld, among other classically $U(1)$ conserving theories.

Observing the redundancy of parameter contributions, it is natural to wonder if one could similarly consider a reduced subset of operators rather than the full set of what is allowed via symmetric and adjoint double copy as described in \eqn{HDPhotonAmp}.  For example, one could instead choose to replace the symmetric color factors, $c^{\text{dd}}_g$, with NLSM symmetric numerators, where $n^{\pi,\text{dd}}_s = s^2$. This choice constitutes a Born-Infeld-like double copy between our higher derivative vector amplitude, $\mathcal{A}_4^{\text{vec}+\text{HD}}$, and NLSM amplitudes, $\mathcal{A}_4^{\text{NLSM}}$. We can describe this numerator level double copy, where $c^{\text{ff/dd}}_g\rightarrow n^{\pi,\text{ff/dd}}_g$, as a generalized product $\otimes$ between theories,
\begin{equation}\label{BIHD}
\mathcal{M}^{\text{BI}+\text{HD}}_4= \mathcal{A}_4^{\text{vec}+\text{HD}}\otimes \mathcal{A}^{\text{NLSM}}_4\,.
\end{equation}
One can see that $\mathcal{M}^{\text{BI}+\text{HD}}_4$ misses some of the available local operators in \eqn{HDPhotonAmp}. However, the freedom of eqns.~(\ref{pppmCT2loop}), (\ref{ppppCT2loop}), and (\ref{finalCs}), even under a shift of available $a_{(x,y)}$ coefficients, suggests that the Born-Infeld operator construction alone should still be sufficient to cancel the anomalous matrix elements through two-loops, a fact that is easy enough to verify explicitly. 
 
Before concluding, it is worth noting that each of the two-loop matrix elements in \eqn{pppmCT2loop} and \eqn{ppppCT2loop} are interesting in their own right. The all-plus counterterms have the potential to further probe tensor structures that can only be encoded locally with a non-adjoint symmetric double copy (parameterized by the Wilson coeffient $c_{(0,3)}^+$). Meanwhile, the second term in the all-plus counterterm and the one-minus contribution, is intriguing for other reasons. This represents yet another opportunity for color-dual $F^3$  to be required for taming the $U(1)$ anomaly in a duality invariant theory \cite{Carrasco:2013ypa,Bern:2017rjw}. Recent studies \cite{Carrasco:2022lbm, Carrasco:2022sck} have shown that double-copy consistent theories of $\text{YM}+F^3$ require an infinite tower of four-point contacts. Whether these contacts are needed for anomaly cancellation at higher-loop order remains an open question. In the case of Born-Infeld theory, the first potential appearance of one of these additional contacts required by double-copy consistency would occur at three-loop.

\section{Discussion}\label{discussion}
In summary, we have shown that we can span the space of four-point photon effective operators using a local-double copy construction. To do so, in Sec.~\ref{symmDoubleCopy} we expressed NLSM amplitudes in a color-dual basis of symmetric color factors, $d^{abc}$. In this form, we found that NLSM amplitudes are color-dual in terms of both adjoint, and non-adjoint local numerators.  In Sec.~\ref{nonAdjNonLocal} we clarify how certain adjoint-type numerators without sufficient higher-derivative support will correspond to amplitudes in theories with high-spin massless particle exchange. This motivated the study of gauge theory operators with parallel symmetric structure, for the purpose of using them in a double-copy construction involving amplitudes in physical theories.

We emphasize the utility of these constructions by sketching an interesting application of these novel color-dual operators in Sec.~\ref{sect:four} to identify anomaly cancelling matrix elements required to preserve the $U(1)$ duality in a quantum Born-Infeld theory. To span all photonic operators, we only need a gauge theory $\text{YM}+F^3$ adjoint structure and a pair of symmetric color-dual building blocks, $f_{ij}f_{kl}$ and $f_{ijkl}$, along with composition with scalar kinematics. While we have chosen to focus on photon operators in this paper, the potential for future studies is nontrivial. 
\paragraph*{Gravity counterterms:} As noted in Sec.~\ref{nonAdjNonLocal}, these symmetric vector building blocks can also be encoded as non-local numerators that obey adjoint kinematics, as long as the spurious poles are cancelled in the double-copy. However, if we were to double copy the adjoint vector numerator in \eqn{nonLocaladjoint} with itself, one would obtain a gravity amplitude with intermediate higher spin modes~\cite{Pavao:2022kog}:
\begin{equation}
    \mathcal{O}^{\text{ff}}_{\text{GR}} = \sum_{g\in \Gamma^{(3)}} \frac{n^{\text{vec,ff}}_g n^{\text{vec,ff}}_g}{d_g} = \frac{(\mathcal{T} \sigma_2^2)^2}{stu}\,.
\end{equation}
This is clearly an unphysical operator. Alternatively we could perform the double-copy with the physical symmetric numerators, $n^{\text{vec,dd}}_g$, which were used to construct the non-local adjoint numerator of \eqn{nonLocaladjoint}, $n^{\text{vec,ff}}_s=n^{\text{vec,dd}}_t-n^{\text{vec,dd}}_u$ used above. This would yield the following gravitational contact:
\begin{equation}
    \mathcal{O}^{\text{dd}}_{\text{GR}} =\sum_{g\in \Gamma^{(3)}} \frac{n^{\text{vec,dd}}_g n^{\text{vec,dd}}_g}{d_g} = \mathcal{T}^2 \sigma_3 \sigma_2\,.
\end{equation}
With our spanning set of symmetric vector building blocks, in conjunction with those identified in \cite{Carrasco:2019yyn} for adjoint kinematics, constructing all physical four-graviton effective operators via a double-copy of local operators is now within reach. 
\paragraph*{Mixing $d^{abc}$ and $f^{abc}$:}
As is clear from the candidate Lagrangian in \eqn{LagrangianSymVec} that gives rise to amplitudes specified by symmetric color dual numerators of \eqn{nonLocaladjoint}, higher multiplcity amplitudes will start mixing $d^{abc}$ and $f^{abc}$ structure constants. Combinations of these color factors obey additional jacobi identities of the form:
\begin{equation}
\label{ddfIdentity}
    f^{a_1a_2e}d^{ea_3a_4}+f^{a_1a_3e}d^{ea_4a_2}+f^{a_1a_4e}d^{ea_2a_3}=0
\end{equation}
These relations follow from Fierz contractions of $f^{abc}$ and $d^{abc}$ in terms of color traces.  In the context of eventual composition to adjoint building blocks, color-dual kinematic building blocks obeying these relations have been explored at five points in~\cite{Carrasco:2021ptp}. Higher-multiplicity and loop-level structure must necessarily account for these additional algebraic constraints for relevant theories to remain color dual.  Mapping the kinematic space that obeys these constraints at higher multiplicity is important future work.
\paragraph*{Double Copy Consistency:}
Once non-adjoint kinematic building blocks are found at higher multiplicity, understanding how they are constrained by factorization will be a critical next step. Recently it was shown \cite{Carrasco:2022lbm} that the many of the Wilson coefficients associated with the adjoint four-point contacts of \cite{Carrasco:2019yyn} are constrained if one demands consistent five-point factorization to a $\text{YM}+F^3$ three-point vertex, and likewise in \cite{Carrasco:2022sck} for $\text{NLSM}+\text{YM}$ theory. This so-called double-copy consistency between the tree-level amplitudes of a color-dual theory could have interesting implications for the double-copy constructed counterterms necessary for BI anomaly cancellation in this paper.
\paragraph*{New Amplitude Relations?} It is enticing to consider what this might mean for further color-dual compression in the landscape of gauge theory amplitudes. For gauge theories that permit adjoint $(f^{abc})^{n-2}$ relations between their $n$-point kinematic weights, the space of local amplitudes reduces from $(n-2)!$ to $(n-3)!$ building blocks. If amplitudes can be equivalently constructed in non-adjoint color-dual forms, then the added kinematic relations could in principle lead to presently unknown amplitude-level redundancy. It is worth noting that for gauge theories with symmetric color-dual numerators at four-point, the algebraic constraint in \eqn{ddfIdentity} first becomes relevant at five-point, where the size of the BCJ basis is two -- inviting a search for additional redundancy. 
\paragraph*{Acknowledgements:}  We would like to thank Alex Edison, James Mangan, Frank Petriello, Laurentiu Rodina, Radu Roiban, Aslan Seifi, and Suna Zekioglu for helpful discussions, related collaboration, and encouragement along  various stages of this project. We are especially grateful to Alex Edison for helpful feedback on earlier versions of this manuscript. This work was supported by the DOE under contract DE-SC0015910 and by the Alfred P. Sloan Foundation. NHP acknowledges the Northwestern University Amplitudes and Insight group, the Department of Physics and Astronomy, and Weinberg College for support.
\bibliography{Refs_symmetricVirtues.bib}

\begin{thebibliography}{48}%
\makeatletter
\providecommand \@ifxundefined [1]{%
 \@ifx{#1\undefined}
}%
\providecommand \@ifnum [1]{%
 \ifnum #1\expandafter \@firstoftwo
 \else \expandafter \@secondoftwo
 \fi
}%
\providecommand \@ifx [1]{%
 \ifx #1\expandafter \@firstoftwo
 \else \expandafter \@secondoftwo
 \fi
}%
\providecommand \natexlab [1]{#1}%
\providecommand \enquote  [1]{``#1''}%
\providecommand \bibnamefont  [1]{#1}%
\providecommand \bibfnamefont [1]{#1}%
\providecommand \citenamefont [1]{#1}%
\providecommand \href@noop [0]{\@secondoftwo}%
\providecommand \href [0]{\begingroup \@sanitize@url \@href}%
\providecommand \@href[1]{\@@startlink{#1}\@@href}%
\providecommand \@@href[1]{\endgroup#1\@@endlink}%
\providecommand \@sanitize@url [0]{\catcode `\\12\catcode `\$12\catcode
  `\&12\catcode `\#12\catcode `\^12\catcode `\_12\catcode `\%12\relax}%
\providecommand \@@startlink[1]{}%
\providecommand \@@endlink[0]{}%
\providecommand \url  [0]{\begingroup\@sanitize@url \@url }%
\providecommand \@url [1]{\endgroup\@href {#1}{\urlprefix }}%
\providecommand \urlprefix  [0]{URL }%
\providecommand \Eprint [0]{\href }%
\providecommand \doibase [0]{http://dx.doi.org/}%
\providecommand \selectlanguage [0]{\@gobble}%
\providecommand \bibinfo  [0]{\@secondoftwo}%
\providecommand \bibfield  [0]{\@secondoftwo}%
\providecommand \translation [1]{[#1]}%
\providecommand \BibitemOpen [0]{}%
\providecommand \bibitemStop [0]{}%
\providecommand \bibitemNoStop [0]{.\EOS\space}%
\providecommand \EOS [0]{\spacefactor3000\relax}%
\providecommand \BibitemShut  [1]{\csname bibitem#1\endcsname}%
\let\auto@bib@innerbib\@empty
\bibitem [{\citenamefont {Georgi}(1993)}]{Georgi:1993mps}%
  \BibitemOpen
  \bibfield  {author} {\bibinfo {author} {\bibfnamefont {H.}~\bibnamefont
  {Georgi}},\ }\bibfield  {title} {\enquote {\bibinfo {title} {{Effective field
  theory}},}\ }\href {\doibase 10.1146/annurev.ns.43.120193.001233} {\bibfield
  {journal} {\bibinfo  {journal} {Ann. Rev. Nucl. Part. Sci.}\ }\textbf
  {\bibinfo {volume} {43}},\ \bibinfo {pages} {209--252} (\bibinfo {year}
  {1993})}\BibitemShut {NoStop}%
\bibitem [{\citenamefont {Weinberg}(1979)}]{Weinberg:1979sa}%
  \BibitemOpen
  \bibfield  {author} {\bibinfo {author} {\bibfnamefont {Steven}\ \bibnamefont
  {Weinberg}},\ }\bibfield  {title} {\enquote {\bibinfo {title} {{Baryon and
  Lepton Nonconserving Processes}},}\ }\href {\doibase
  10.1103/PhysRevLett.43.1566} {\bibfield  {journal} {\bibinfo  {journal}
  {Phys. Rev. Lett.}\ }\textbf {\bibinfo {volume} {43}},\ \bibinfo {pages}
  {1566--1570} (\bibinfo {year} {1979})}\BibitemShut {NoStop}%
\bibitem [{\citenamefont {Buchmuller}\ and\ \citenamefont
  {Wyler}(1986)}]{Buchmuller:1985jz}%
  \BibitemOpen
  \bibfield  {author} {\bibinfo {author} {\bibfnamefont {W.}~\bibnamefont
  {Buchmuller}}\ and\ \bibinfo {author} {\bibfnamefont {D.}~\bibnamefont
  {Wyler}},\ }\bibfield  {title} {\enquote {\bibinfo {title} {{Effective
  Lagrangian Analysis of New Interactions and Flavor Conservation}},}\ }\href
  {\doibase 10.1016/0550-3213(86)90262-2} {\bibfield  {journal} {\bibinfo
  {journal} {Nucl. Phys. B}\ }\textbf {\bibinfo {volume} {268}},\ \bibinfo
  {pages} {621--653} (\bibinfo {year} {1986})}\BibitemShut {NoStop}%
\bibitem [{\citenamefont {Isgur}\ and\ \citenamefont
  {Wise}(1989)}]{Isgur:1989vq}%
  \BibitemOpen
  \bibfield  {author} {\bibinfo {author} {\bibfnamefont {Nathan}\ \bibnamefont
  {Isgur}}\ and\ \bibinfo {author} {\bibfnamefont {Mark~B.}\ \bibnamefont
  {Wise}},\ }\bibfield  {title} {\enquote {\bibinfo {title} {{Weak Decays of
  Heavy Mesons in the Static Quark Approximation}},}\ }\href {\doibase
  10.1016/0370-2693(89)90566-2} {\bibfield  {journal} {\bibinfo  {journal}
  {Phys. Lett. B}\ }\textbf {\bibinfo {volume} {232}},\ \bibinfo {pages}
  {113--117} (\bibinfo {year} {1989})}\BibitemShut {NoStop}%
\bibitem [{\citenamefont {Bauer}\ \emph {et~al.}(2002)\citenamefont {Bauer},
  \citenamefont {Fleming}, \citenamefont {Pirjol}, \citenamefont {Rothstein},\
  and\ \citenamefont {Stewart}}]{Bauer:2002nz}%
  \BibitemOpen
  \bibfield  {author} {\bibinfo {author} {\bibfnamefont {Christian~W.}\
  \bibnamefont {Bauer}}, \bibinfo {author} {\bibfnamefont {Sean}\ \bibnamefont
  {Fleming}}, \bibinfo {author} {\bibfnamefont {Dan}\ \bibnamefont {Pirjol}},
  \bibinfo {author} {\bibfnamefont {Ira~Z.}\ \bibnamefont {Rothstein}}, \ and\
  \bibinfo {author} {\bibfnamefont {Iain~W.}\ \bibnamefont {Stewart}},\
  }\bibfield  {title} {\enquote {\bibinfo {title} {{Hard scattering
  factorization from effective field theory}},}\ }\href {\doibase
  10.1103/PhysRevD.66.014017} {\bibfield  {journal} {\bibinfo  {journal} {Phys.
  Rev. D}\ }\textbf {\bibinfo {volume} {66}},\ \bibinfo {pages} {014017}
  (\bibinfo {year} {2002})},\ \Eprint {http://arxiv.org/abs/hep-ph/0202088}
  {arXiv:hep-ph/0202088} \BibitemShut {NoStop}%
\bibitem [{\citenamefont {Brivio}\ and\ \citenamefont
  {Trott}(2019)}]{Brivio:2017vri}%
  \BibitemOpen
  \bibfield  {author} {\bibinfo {author} {\bibfnamefont {Ilaria}\ \bibnamefont
  {Brivio}}\ and\ \bibinfo {author} {\bibfnamefont {Michael}\ \bibnamefont
  {Trott}},\ }\bibfield  {title} {\enquote {\bibinfo {title} {{The Standard
  Model as an Effective Field Theory}},}\ }\href {\doibase
  10.1016/j.physrep.2018.11.002} {\bibfield  {journal} {\bibinfo  {journal}
  {Phys. Rept.}\ }\textbf {\bibinfo {volume} {793}},\ \bibinfo {pages} {1--98}
  (\bibinfo {year} {2019})},\ \Eprint {http://arxiv.org/abs/1706.08945}
  {arXiv:1706.08945 [hep-ph]} \BibitemShut {NoStop}%
\bibitem [{\citenamefont {Goldberger}\ and\ \citenamefont
  {Rothstein}(2006)}]{Goldberger:2004jt}%
  \BibitemOpen
  \bibfield  {author} {\bibinfo {author} {\bibfnamefont {Walter~D.}\
  \bibnamefont {Goldberger}}\ and\ \bibinfo {author} {\bibfnamefont {Ira~Z.}\
  \bibnamefont {Rothstein}},\ }\bibfield  {title} {\enquote {\bibinfo {title}
  {{An Effective field theory of gravity for extended objects}},}\ }\href
  {\doibase 10.1103/PhysRevD.73.104029} {\bibfield  {journal} {\bibinfo
  {journal} {Phys. Rev. D}\ }\textbf {\bibinfo {volume} {73}},\ \bibinfo
  {pages} {104029} (\bibinfo {year} {2006})},\ \Eprint
  {http://arxiv.org/abs/hep-th/0409156} {arXiv:hep-th/0409156} \BibitemShut
  {NoStop}%
\bibitem [{\citenamefont {Porto}(2006)}]{Porto:2005ac}%
  \BibitemOpen
  \bibfield  {author} {\bibinfo {author} {\bibfnamefont {Rafael~A.}\
  \bibnamefont {Porto}},\ }\bibfield  {title} {\enquote {\bibinfo {title}
  {{Post-Newtonian corrections to the motion of spinning bodies in NRGR}},}\
  }\href {\doibase 10.1103/PhysRevD.73.104031} {\bibfield  {journal} {\bibinfo
  {journal} {Phys. Rev. D}\ }\textbf {\bibinfo {volume} {73}},\ \bibinfo
  {pages} {104031} (\bibinfo {year} {2006})},\ \Eprint
  {http://arxiv.org/abs/gr-qc/0511061} {arXiv:gr-qc/0511061} \BibitemShut
  {NoStop}%
\bibitem [{\citenamefont {Cheung}\ \emph {et~al.}(2018)\citenamefont {Cheung},
  \citenamefont {Rothstein},\ and\ \citenamefont {Solon}}]{Cheung:2018wkq}%
  \BibitemOpen
  \bibfield  {author} {\bibinfo {author} {\bibfnamefont {Clifford}\
  \bibnamefont {Cheung}}, \bibinfo {author} {\bibfnamefont {Ira~Z.}\
  \bibnamefont {Rothstein}}, \ and\ \bibinfo {author} {\bibfnamefont
  {Mikhail~P.}\ \bibnamefont {Solon}},\ }\bibfield  {title} {\enquote {\bibinfo
  {title} {{From Scattering Amplitudes to Classical Potentials in the
  Post-Minkowskian Expansion}},}\ }\href {\doibase
  10.1103/PhysRevLett.121.251101} {\bibfield  {journal} {\bibinfo  {journal}
  {Phys. Rev. Lett.}\ }\textbf {\bibinfo {volume} {121}},\ \bibinfo {pages}
  {251101} (\bibinfo {year} {2018})},\ \Eprint
  {http://arxiv.org/abs/1808.02489} {arXiv:1808.02489 [hep-th]} \BibitemShut
  {NoStop}%
\bibitem [{\citenamefont {Edison}\ and\ \citenamefont
  {Levi}(2023)}]{Edison:2022cdu}%
  \BibitemOpen
  \bibfield  {author} {\bibinfo {author} {\bibfnamefont {Alex}\ \bibnamefont
  {Edison}}\ and\ \bibinfo {author} {\bibfnamefont {Mich\`ele}\ \bibnamefont
  {Levi}},\ }\bibfield  {title} {\enquote {\bibinfo {title} {{A tale of tails
  through generalized unitarity}},}\ }\href {\doibase
  10.1016/j.physletb.2022.137634} {\bibfield  {journal} {\bibinfo  {journal}
  {Phys. Lett. B}\ }\textbf {\bibinfo {volume} {837}},\ \bibinfo {pages}
  {137634} (\bibinfo {year} {2023})},\ \Eprint
  {http://arxiv.org/abs/2202.04674} {arXiv:2202.04674 [hep-th]} \BibitemShut
  {NoStop}%
\bibitem [{\citenamefont {Creminelli}\ \emph {et~al.}(2006)\citenamefont
  {Creminelli}, \citenamefont {Luty}, \citenamefont {Nicolis},\ and\
  \citenamefont {Senatore}}]{Creminelli:2006xe}%
  \BibitemOpen
  \bibfield  {author} {\bibinfo {author} {\bibfnamefont {Paolo}\ \bibnamefont
  {Creminelli}}, \bibinfo {author} {\bibfnamefont {Markus~A.}\ \bibnamefont
  {Luty}}, \bibinfo {author} {\bibfnamefont {Alberto}\ \bibnamefont {Nicolis}},
  \ and\ \bibinfo {author} {\bibfnamefont {Leonardo}\ \bibnamefont
  {Senatore}},\ }\bibfield  {title} {\enquote {\bibinfo {title} {{Starting the
  Universe: Stable Violation of the Null Energy Condition and Non-standard
  Cosmologies}},}\ }\href {\doibase 10.1088/1126-6708/2006/12/080} {\bibfield
  {journal} {\bibinfo  {journal} {JHEP}\ }\textbf {\bibinfo {volume} {12}},\
  \bibinfo {pages} {080} (\bibinfo {year} {2006})},\ \Eprint
  {http://arxiv.org/abs/hep-th/0606090} {arXiv:hep-th/0606090} \BibitemShut
  {NoStop}%
\bibitem [{\citenamefont {Cheung}\ \emph {et~al.}(2008)\citenamefont {Cheung},
  \citenamefont {Creminelli}, \citenamefont {Fitzpatrick}, \citenamefont
  {Kaplan},\ and\ \citenamefont {Senatore}}]{Cheung:2007st}%
  \BibitemOpen
  \bibfield  {author} {\bibinfo {author} {\bibfnamefont {Clifford}\
  \bibnamefont {Cheung}}, \bibinfo {author} {\bibfnamefont {Paolo}\
  \bibnamefont {Creminelli}}, \bibinfo {author} {\bibfnamefont {A.~Liam}\
  \bibnamefont {Fitzpatrick}}, \bibinfo {author} {\bibfnamefont {Jared}\
  \bibnamefont {Kaplan}}, \ and\ \bibinfo {author} {\bibfnamefont {Leonardo}\
  \bibnamefont {Senatore}},\ }\bibfield  {title} {\enquote {\bibinfo {title}
  {{The Effective Field Theory of Inflation}},}\ }\href {\doibase
  10.1088/1126-6708/2008/03/014} {\bibfield  {journal} {\bibinfo  {journal}
  {JHEP}\ }\textbf {\bibinfo {volume} {03}},\ \bibinfo {pages} {014} (\bibinfo
  {year} {2008})},\ \Eprint {http://arxiv.org/abs/0709.0293} {arXiv:0709.0293
  [hep-th]} \BibitemShut {NoStop}%
\bibitem [{\citenamefont {Weinberg}(2008)}]{Weinberg:2008hq}%
  \BibitemOpen
  \bibfield  {author} {\bibinfo {author} {\bibfnamefont {Steven}\ \bibnamefont
  {Weinberg}},\ }\bibfield  {title} {\enquote {\bibinfo {title} {{Effective
  Field Theory for Inflation}},}\ }\href {\doibase 10.1103/PhysRevD.77.123541}
  {\bibfield  {journal} {\bibinfo  {journal} {Phys. Rev. D}\ }\textbf {\bibinfo
  {volume} {77}},\ \bibinfo {pages} {123541} (\bibinfo {year} {2008})},\
  \Eprint {http://arxiv.org/abs/0804.4291} {arXiv:0804.4291 [hep-th]}
  \BibitemShut {NoStop}%
\bibitem [{\citenamefont {Burgess}\ \emph {et~al.}(2009)\citenamefont
  {Burgess}, \citenamefont {Lee},\ and\ \citenamefont
  {Trott}}]{Burgess:2009ea}%
  \BibitemOpen
  \bibfield  {author} {\bibinfo {author} {\bibfnamefont {C.~P.}\ \bibnamefont
  {Burgess}}, \bibinfo {author} {\bibfnamefont {Hyun~Min}\ \bibnamefont {Lee}},
  \ and\ \bibinfo {author} {\bibfnamefont {Michael}\ \bibnamefont {Trott}},\
  }\bibfield  {title} {\enquote {\bibinfo {title} {{Power-counting and the
  Validity of the Classical Approximation During Inflation}},}\ }\href
  {\doibase 10.1088/1126-6708/2009/09/103} {\bibfield  {journal} {\bibinfo
  {journal} {JHEP}\ }\textbf {\bibinfo {volume} {09}},\ \bibinfo {pages} {103}
  (\bibinfo {year} {2009})},\ \Eprint {http://arxiv.org/abs/0902.4465}
  {arXiv:0902.4465 [hep-ph]} \BibitemShut {NoStop}%
\bibitem [{\citenamefont {Gubitosi}\ \emph {et~al.}(2013)\citenamefont
  {Gubitosi}, \citenamefont {Piazza},\ and\ \citenamefont
  {Vernizzi}}]{Gubitosi:2012hu}%
  \BibitemOpen
  \bibfield  {author} {\bibinfo {author} {\bibfnamefont {Giulia}\ \bibnamefont
  {Gubitosi}}, \bibinfo {author} {\bibfnamefont {Federico}\ \bibnamefont
  {Piazza}}, \ and\ \bibinfo {author} {\bibfnamefont {Filippo}\ \bibnamefont
  {Vernizzi}},\ }\bibfield  {title} {\enquote {\bibinfo {title} {{The Effective
  Field Theory of Dark Energy}},}\ }\href {\doibase
  10.1088/1475-7516/2013/02/032} {\bibfield  {journal} {\bibinfo  {journal}
  {JCAP}\ }\textbf {\bibinfo {volume} {02}},\ \bibinfo {pages} {032} (\bibinfo
  {year} {2013})},\ \Eprint {http://arxiv.org/abs/1210.0201} {arXiv:1210.0201
  [hep-th]} \BibitemShut {NoStop}%
\bibitem [{\citenamefont {Baumann}\ \emph {et~al.}(2012)\citenamefont
  {Baumann}, \citenamefont {Nicolis}, \citenamefont {Senatore},\ and\
  \citenamefont {Zaldarriaga}}]{Baumann:2010tm}%
  \BibitemOpen
  \bibfield  {author} {\bibinfo {author} {\bibfnamefont {Daniel}\ \bibnamefont
  {Baumann}}, \bibinfo {author} {\bibfnamefont {Alberto}\ \bibnamefont
  {Nicolis}}, \bibinfo {author} {\bibfnamefont {Leonardo}\ \bibnamefont
  {Senatore}}, \ and\ \bibinfo {author} {\bibfnamefont {Matias}\ \bibnamefont
  {Zaldarriaga}},\ }\bibfield  {title} {\enquote {\bibinfo {title}
  {{Cosmological Non-Linearities as an Effective Fluid}},}\ }\href {\doibase
  10.1088/1475-7516/2012/07/051} {\bibfield  {journal} {\bibinfo  {journal}
  {JCAP}\ }\textbf {\bibinfo {volume} {07}},\ \bibinfo {pages} {051} (\bibinfo
  {year} {2012})},\ \Eprint {http://arxiv.org/abs/1004.2488} {arXiv:1004.2488
  [astro-ph.CO]} \BibitemShut {NoStop}%
\bibitem [{\citenamefont {Carrasco}\ \emph {et~al.}(2012)\citenamefont
  {Carrasco}, \citenamefont {Hertzberg},\ and\ \citenamefont
  {Senatore}}]{Carrasco:2012cv}%
  \BibitemOpen
  \bibfield  {author} {\bibinfo {author} {\bibfnamefont {John Joseph~M.}\
  \bibnamefont {Carrasco}}, \bibinfo {author} {\bibfnamefont {Mark~P.}\
  \bibnamefont {Hertzberg}}, \ and\ \bibinfo {author} {\bibfnamefont
  {Leonardo}\ \bibnamefont {Senatore}},\ }\bibfield  {title} {\enquote
  {\bibinfo {title} {{The Effective Field Theory of Cosmological Large Scale
  Structures}},}\ }\href {\doibase 10.1007/JHEP09(2012)082} {\bibfield
  {journal} {\bibinfo  {journal} {JHEP}\ }\textbf {\bibinfo {volume} {09}},\
  \bibinfo {pages} {082} (\bibinfo {year} {2012})},\ \Eprint
  {http://arxiv.org/abs/1206.2926} {arXiv:1206.2926 [astro-ph.CO]} \BibitemShut
  {NoStop}%
\bibitem [{\citenamefont {Porto}\ \emph {et~al.}(2014)\citenamefont {Porto},
  \citenamefont {Senatore},\ and\ \citenamefont {Zaldarriaga}}]{Porto:2013qua}%
  \BibitemOpen
  \bibfield  {author} {\bibinfo {author} {\bibfnamefont {Rafael~A.}\
  \bibnamefont {Porto}}, \bibinfo {author} {\bibfnamefont {Leonardo}\
  \bibnamefont {Senatore}}, \ and\ \bibinfo {author} {\bibfnamefont {Matias}\
  \bibnamefont {Zaldarriaga}},\ }\bibfield  {title} {\enquote {\bibinfo {title}
  {{The Lagrangian-space Effective Field Theory of Large Scale Structures}},}\
  }\href {\doibase 10.1088/1475-7516/2014/05/022} {\bibfield  {journal}
  {\bibinfo  {journal} {JCAP}\ }\textbf {\bibinfo {volume} {05}},\ \bibinfo
  {pages} {022} (\bibinfo {year} {2014})},\ \Eprint
  {http://arxiv.org/abs/1311.2168} {arXiv:1311.2168 [astro-ph.CO]} \BibitemShut
  {NoStop}%
\bibitem [{\citenamefont {Lehman}\ and\ \citenamefont
  {Martin}(2015)}]{Lehman:2015via}%
  \BibitemOpen
  \bibfield  {author} {\bibinfo {author} {\bibfnamefont {Landon}\ \bibnamefont
  {Lehman}}\ and\ \bibinfo {author} {\bibfnamefont {Adam}\ \bibnamefont
  {Martin}},\ }\bibfield  {title} {\enquote {\bibinfo {title} {{Hilbert Series
  for Constructing Lagrangians: expanding the phenomenologist's toolbox}},}\
  }\href {\doibase 10.1103/PhysRevD.91.105014} {\bibfield  {journal} {\bibinfo
  {journal} {Phys. Rev. D}\ }\textbf {\bibinfo {volume} {91}},\ \bibinfo
  {pages} {105014} (\bibinfo {year} {2015})},\ \Eprint
  {http://arxiv.org/abs/1503.07537} {arXiv:1503.07537 [hep-ph]} \BibitemShut
  {NoStop}%
\bibitem [{\citenamefont {Henning}\ \emph {et~al.}(2016)\citenamefont
  {Henning}, \citenamefont {Lu}, \citenamefont {Melia},\ and\ \citenamefont
  {Murayama}}]{Henning:2015daa}%
  \BibitemOpen
  \bibfield  {author} {\bibinfo {author} {\bibfnamefont {Brian}\ \bibnamefont
  {Henning}}, \bibinfo {author} {\bibfnamefont {Xiaochuan}\ \bibnamefont {Lu}},
  \bibinfo {author} {\bibfnamefont {Tom}\ \bibnamefont {Melia}}, \ and\
  \bibinfo {author} {\bibfnamefont {Hitoshi}\ \bibnamefont {Murayama}},\
  }\bibfield  {title} {\enquote {\bibinfo {title} {{Hilbert series and operator
  bases with derivatives in effective field theories}},}\ }\href {\doibase
  10.1007/s00220-015-2518-2} {\bibfield  {journal} {\bibinfo  {journal}
  {Commun. Math. Phys.}\ }\textbf {\bibinfo {volume} {347}},\ \bibinfo {pages}
  {363--388} (\bibinfo {year} {2016})},\ \Eprint
  {http://arxiv.org/abs/1507.07240} {arXiv:1507.07240 [hep-th]} \BibitemShut
  {NoStop}%
\bibitem [{\citenamefont {Elvang}\ \emph {et~al.}(2019)\citenamefont {Elvang},
  \citenamefont {Hadjiantonis}, \citenamefont {Jones},\ and\ \citenamefont
  {Paranjape}}]{Elvang:2018dco}%
  \BibitemOpen
  \bibfield  {author} {\bibinfo {author} {\bibfnamefont {Henriette}\
  \bibnamefont {Elvang}}, \bibinfo {author} {\bibfnamefont {Marios}\
  \bibnamefont {Hadjiantonis}}, \bibinfo {author} {\bibfnamefont {Callum
  R.~T.}\ \bibnamefont {Jones}}, \ and\ \bibinfo {author} {\bibfnamefont
  {Shruti}\ \bibnamefont {Paranjape}},\ }\bibfield  {title} {\enquote {\bibinfo
  {title} {{Soft Bootstrap and Supersymmetry}},}\ }\href {\doibase
  10.1007/JHEP01(2019)195} {\bibfield  {journal} {\bibinfo  {journal} {JHEP}\
  }\textbf {\bibinfo {volume} {01}},\ \bibinfo {pages} {195} (\bibinfo {year}
  {2019})},\ \Eprint {http://arxiv.org/abs/1806.06079} {arXiv:1806.06079
  [hep-th]} \BibitemShut {NoStop}%
\bibitem [{\citenamefont {Shadmi}\ and\ \citenamefont
  {Weiss}(2019)}]{Shadmi:2018xan}%
  \BibitemOpen
  \bibfield  {author} {\bibinfo {author} {\bibfnamefont {Yael}\ \bibnamefont
  {Shadmi}}\ and\ \bibinfo {author} {\bibfnamefont {Yaniv}\ \bibnamefont
  {Weiss}},\ }\bibfield  {title} {\enquote {\bibinfo {title} {{Effective Field
  Theory Amplitudes the On-Shell Way: Scalar and Vector Couplings to
  Gluons}},}\ }\href {\doibase 10.1007/JHEP02(2019)165} {\bibfield  {journal}
  {\bibinfo  {journal} {JHEP}\ }\textbf {\bibinfo {volume} {02}},\ \bibinfo
  {pages} {165} (\bibinfo {year} {2019})},\ \Eprint
  {http://arxiv.org/abs/1809.09644} {arXiv:1809.09644 [hep-ph]} \BibitemShut
  {NoStop}%
\bibitem [{\citenamefont {Durieux}\ \emph {et~al.}(2020)\citenamefont
  {Durieux}, \citenamefont {Kitahara}, \citenamefont {Shadmi},\ and\
  \citenamefont {Weiss}}]{Durieux:2019eor}%
  \BibitemOpen
  \bibfield  {author} {\bibinfo {author} {\bibfnamefont {Gauthier}\
  \bibnamefont {Durieux}}, \bibinfo {author} {\bibfnamefont {Teppei}\
  \bibnamefont {Kitahara}}, \bibinfo {author} {\bibfnamefont {Yael}\
  \bibnamefont {Shadmi}}, \ and\ \bibinfo {author} {\bibfnamefont {Yaniv}\
  \bibnamefont {Weiss}},\ }\bibfield  {title} {\enquote {\bibinfo {title} {{The
  electroweak effective field theory from on-shell amplitudes}},}\ }\href
  {\doibase 10.1007/JHEP01(2020)119} {\bibfield  {journal} {\bibinfo  {journal}
  {JHEP}\ }\textbf {\bibinfo {volume} {01}},\ \bibinfo {pages} {119} (\bibinfo
  {year} {2020})},\ \Eprint {http://arxiv.org/abs/1909.10551} {arXiv:1909.10551
  [hep-ph]} \BibitemShut {NoStop}%
\bibitem [{\citenamefont {Arkani-Hamed}\ \emph {et~al.}(2021)\citenamefont
  {Arkani-Hamed}, \citenamefont {Huang},\ and\ \citenamefont
  {Huang}}]{Arkani-Hamed:2020blm}%
  \BibitemOpen
  \bibfield  {author} {\bibinfo {author} {\bibfnamefont {Nima}\ \bibnamefont
  {Arkani-Hamed}}, \bibinfo {author} {\bibfnamefont {Tzu-Chen}\ \bibnamefont
  {Huang}}, \ and\ \bibinfo {author} {\bibfnamefont {Yu-Tin}\ \bibnamefont
  {Huang}},\ }\bibfield  {title} {\enquote {\bibinfo {title} {{The
  EFT-Hedron}},}\ }\href {\doibase 10.1007/JHEP05(2021)259} {\bibfield
  {journal} {\bibinfo  {journal} {JHEP}\ }\textbf {\bibinfo {volume} {05}},\
  \bibinfo {pages} {259} (\bibinfo {year} {2021})},\ \Eprint
  {http://arxiv.org/abs/2012.15849} {arXiv:2012.15849 [hep-th]} \BibitemShut
  {NoStop}%
\bibitem [{\citenamefont {Caron-Huot}\ \emph {et~al.}(2021)\citenamefont
  {Caron-Huot}, \citenamefont {Mazac}, \citenamefont {Rastelli},\ and\
  \citenamefont {Simmons-Duffin}}]{Caron-Huot:2021rmr}%
  \BibitemOpen
  \bibfield  {author} {\bibinfo {author} {\bibfnamefont {Simon}\ \bibnamefont
  {Caron-Huot}}, \bibinfo {author} {\bibfnamefont {Dalimil}\ \bibnamefont
  {Mazac}}, \bibinfo {author} {\bibfnamefont {Leonardo}\ \bibnamefont
  {Rastelli}}, \ and\ \bibinfo {author} {\bibfnamefont {David}\ \bibnamefont
  {Simmons-Duffin}},\ }\bibfield  {title} {\enquote {\bibinfo {title} {{Sharp
  boundaries for the swampland}},}\ }\href {\doibase 10.1007/JHEP07(2021)110}
  {\bibfield  {journal} {\bibinfo  {journal} {JHEP}\ }\textbf {\bibinfo
  {volume} {07}},\ \bibinfo {pages} {110} (\bibinfo {year} {2021})},\ \Eprint
  {http://arxiv.org/abs/2102.08951} {arXiv:2102.08951 [hep-th]} \BibitemShut
  {NoStop}%
\bibitem [{\citenamefont {Bern}\ \emph {et~al.}(2021)\citenamefont {Bern},
  \citenamefont {Kosmopoulos},\ and\ \citenamefont {Zhiboedov}}]{Bern:2021ppb}%
  \BibitemOpen
  \bibfield  {author} {\bibinfo {author} {\bibfnamefont {Zvi}\ \bibnamefont
  {Bern}}, \bibinfo {author} {\bibfnamefont {Dimitrios}\ \bibnamefont
  {Kosmopoulos}}, \ and\ \bibinfo {author} {\bibfnamefont {Alexander}\
  \bibnamefont {Zhiboedov}},\ }\bibfield  {title} {\enquote {\bibinfo {title}
  {{Gravitational effective field theory islands, low-spin dominance, and the
  four-graviton amplitude}},}\ }\href {\doibase 10.1088/1751-8121/ac0e51}
  {\bibfield  {journal} {\bibinfo  {journal} {J. Phys. A}\ }\textbf {\bibinfo
  {volume} {54}},\ \bibinfo {pages} {344002} (\bibinfo {year} {2021})},\
  \Eprint {http://arxiv.org/abs/2103.12728} {arXiv:2103.12728 [hep-th]}
  \BibitemShut {NoStop}%
\bibitem [{\citenamefont {Bern}\ \emph
  {et~al.}(2022{\natexlab{a}})\citenamefont {Bern}, \citenamefont {Herrmann},
  \citenamefont {Kosmopoulos},\ and\ \citenamefont {Roiban}}]{Bern:2022yes}%
  \BibitemOpen
  \bibfield  {author} {\bibinfo {author} {\bibfnamefont {Zvi}\ \bibnamefont
  {Bern}}, \bibinfo {author} {\bibfnamefont {Enrico}\ \bibnamefont {Herrmann}},
  \bibinfo {author} {\bibfnamefont {Dimitrios}\ \bibnamefont {Kosmopoulos}}, \
  and\ \bibinfo {author} {\bibfnamefont {Radu}\ \bibnamefont {Roiban}},\
  }\bibfield  {title} {\enquote {\bibinfo {title} {{Effective Field Theory
  Islands from Perturbative and Nonperturbative Four-Graviton Amplitudes}},}\
  }\href@noop {} {\  (\bibinfo {year} {2022}{\natexlab{a}})},\ \Eprint
  {http://arxiv.org/abs/2205.01655} {arXiv:2205.01655 [hep-th]} \BibitemShut
  {NoStop}%
\bibitem [{\citenamefont {Bern}\ \emph {et~al.}(2008)\citenamefont {Bern},
  \citenamefont {Carrasco},\ and\ \citenamefont {Johansson}}]{BCJ}%
  \BibitemOpen
  \bibfield  {author} {\bibinfo {author} {\bibfnamefont {Z.}~\bibnamefont
  {Bern}}, \bibinfo {author} {\bibfnamefont {J.~J.~M.}\ \bibnamefont
  {Carrasco}}, \ and\ \bibinfo {author} {\bibfnamefont {Henrik}\ \bibnamefont
  {Johansson}},\ }\bibfield  {title} {\enquote {\bibinfo {title} {{New
  relations for gauge-theory amplitudes}},}\ }\href {\doibase
  10.1103/PhysRevD.78.085011} {\bibfield  {journal} {\bibinfo  {journal} {Phys.
  Rev.}\ }\textbf {\bibinfo {volume} {D78}},\ \bibinfo {pages} {085011}
  (\bibinfo {year} {2008})},\ \Eprint {http://arxiv.org/abs/0805.3993}
  {arXiv:0805.3993 [hep-ph]} \BibitemShut {NoStop}%
\bibitem [{\citenamefont {Bern}\ \emph
  {et~al.}(2010{\natexlab{a}})\citenamefont {Bern}, \citenamefont {Carrasco},\
  and\ \citenamefont {Johansson}}]{BCJLoop}%
  \BibitemOpen
  \bibfield  {author} {\bibinfo {author} {\bibfnamefont {Zvi}\ \bibnamefont
  {Bern}}, \bibinfo {author} {\bibfnamefont {John Joseph~M.}\ \bibnamefont
  {Carrasco}}, \ and\ \bibinfo {author} {\bibfnamefont {Henrik}\ \bibnamefont
  {Johansson}},\ }\bibfield  {title} {\enquote {\bibinfo {title} {{Perturbative
  quantum gravity as a double copy of gauge theory}},}\ }\href {\doibase
  10.1103/PhysRevLett.105.061602} {\bibfield  {journal} {\bibinfo  {journal}
  {Phys. Rev. Lett.}\ }\textbf {\bibinfo {volume} {105}},\ \bibinfo {pages}
  {061602} (\bibinfo {year} {2010}{\natexlab{a}})},\ \Eprint
  {http://arxiv.org/abs/1004.0476} {arXiv:1004.0476 [hep-th]} \BibitemShut
  {NoStop}%
\bibitem [{\citenamefont {Bern}\ \emph
  {et~al.}(2017{\natexlab{a}})\citenamefont {Bern}, \citenamefont {Chi},
  \citenamefont {Dixon},\ and\ \citenamefont {Edison}}]{Bern:2017puu}%
  \BibitemOpen
  \bibfield  {author} {\bibinfo {author} {\bibfnamefont {Zvi}\ \bibnamefont
  {Bern}}, \bibinfo {author} {\bibfnamefont {Huan-Hang}\ \bibnamefont {Chi}},
  \bibinfo {author} {\bibfnamefont {Lance}\ \bibnamefont {Dixon}}, \ and\
  \bibinfo {author} {\bibfnamefont {Alex}\ \bibnamefont {Edison}},\ }\bibfield
  {title} {\enquote {\bibinfo {title} {{Two-loop renormalization of quantum
  gravity simplified}},}\ }\href {\doibase 10.1103/PhysRevD.95.046013}
  {\bibfield  {journal} {\bibinfo  {journal} {Phys. Rev.}\ }\textbf {\bibinfo
  {volume} {D95}},\ \bibinfo {pages} {046013} (\bibinfo {year}
  {2017}{\natexlab{a}})},\ \Eprint {http://arxiv.org/abs/1701.02422}
  {arXiv:1701.02422 [hep-th]} \BibitemShut {NoStop}%
\bibitem [{\citenamefont {Bern}\ \emph
  {et~al.}(2017{\natexlab{b}})\citenamefont {Bern}, \citenamefont {Edison},
  \citenamefont {Kosower},\ and\ \citenamefont
  {Parra-Martinez}}]{Bern:2017tuc}%
  \BibitemOpen
  \bibfield  {author} {\bibinfo {author} {\bibfnamefont {Zvi}\ \bibnamefont
  {Bern}}, \bibinfo {author} {\bibfnamefont {Alex}\ \bibnamefont {Edison}},
  \bibinfo {author} {\bibfnamefont {David}\ \bibnamefont {Kosower}}, \ and\
  \bibinfo {author} {\bibfnamefont {Julio}\ \bibnamefont {Parra-Martinez}},\
  }\bibfield  {title} {\enquote {\bibinfo {title} {{Curvature-squared
  multiplets, evanescent effects, and the U(1) anomaly in ${\cal N}=4$
  supergravity}},}\ }\href {\doibase 10.1103/PhysRevD.96.066004} {\bibfield
  {journal} {\bibinfo  {journal} {Phys. Rev.}\ }\textbf {\bibinfo {volume}
  {D96}},\ \bibinfo {pages} {066004} (\bibinfo {year} {2017}{\natexlab{b}})},\
  \Eprint {http://arxiv.org/abs/1706.01486} {arXiv:1706.01486 [hep-th]}
  \BibitemShut {NoStop}%
\bibitem [{\citenamefont {Carrasco}\ \emph {et~al.}(2020)\citenamefont
  {Carrasco}, \citenamefont {Rodina}, \citenamefont {Yin},\ and\ \citenamefont
  {Zekioglu}}]{Carrasco:2019yyn}%
  \BibitemOpen
  \bibfield  {author} {\bibinfo {author} {\bibfnamefont {John Joseph~M.}\
  \bibnamefont {Carrasco}}, \bibinfo {author} {\bibfnamefont {Laurentiu}\
  \bibnamefont {Rodina}}, \bibinfo {author} {\bibfnamefont {Zanpeng}\
  \bibnamefont {Yin}}, \ and\ \bibinfo {author} {\bibfnamefont {Suna}\
  \bibnamefont {Zekioglu}},\ }\bibfield  {title} {\enquote {\bibinfo {title}
  {{Simple encoding of higher derivative gauge and gravity counterterms}},}\
  }\href {\doibase 10.1103/PhysRevLett.125.251602} {\bibfield  {journal}
  {\bibinfo  {journal} {Phys. Rev. Lett.}\ }\textbf {\bibinfo {volume} {125}},\
  \bibinfo {pages} {251602} (\bibinfo {year} {2020})},\ \Eprint
  {http://arxiv.org/abs/1910.12850} {arXiv:1910.12850 [hep-th]} \BibitemShut
  {NoStop}%
\bibitem [{\citenamefont {Low}\ and\ \citenamefont {Yin}(2020)}]{Low:2019wuv}%
  \BibitemOpen
  \bibfield  {author} {\bibinfo {author} {\bibfnamefont {Ian}\ \bibnamefont
  {Low}}\ and\ \bibinfo {author} {\bibfnamefont {Zhewei}\ \bibnamefont {Yin}},\
  }\bibfield  {title} {\enquote {\bibinfo {title} {{New Flavor-Kinematics
  Dualities and Extensions of Nonlinear Sigma Models}},}\ }\href {\doibase
  10.1016/j.physletb.2020.135544} {\bibfield  {journal} {\bibinfo  {journal}
  {Phys. Lett. B}\ }\textbf {\bibinfo {volume} {807}},\ \bibinfo {pages}
  {135544} (\bibinfo {year} {2020})},\ \Eprint
  {http://arxiv.org/abs/1911.08490} {arXiv:1911.08490 [hep-th]} \BibitemShut
  {NoStop}%
\bibitem [{\citenamefont {Low}\ \emph {et~al.}(2021)\citenamefont {Low},
  \citenamefont {Rodina},\ and\ \citenamefont {Yin}}]{Low:2020ubn}%
  \BibitemOpen
  \bibfield  {author} {\bibinfo {author} {\bibfnamefont {Ian}\ \bibnamefont
  {Low}}, \bibinfo {author} {\bibfnamefont {Laurentiu}\ \bibnamefont {Rodina}},
  \ and\ \bibinfo {author} {\bibfnamefont {Zhewei}\ \bibnamefont {Yin}},\
  }\bibfield  {title} {\enquote {\bibinfo {title} {{Double Copy in Higher
  Derivative Operators of Nambu-Goldstone Bosons}},}\ }\href {\doibase
  10.1103/PhysRevD.103.025004} {\bibfield  {journal} {\bibinfo  {journal}
  {Phys. Rev. D}\ }\textbf {\bibinfo {volume} {103}},\ \bibinfo {pages}
  {025004} (\bibinfo {year} {2021})},\ \Eprint
  {http://arxiv.org/abs/2009.00008} {arXiv:2009.00008 [hep-th]} \BibitemShut
  {NoStop}%
\bibitem [{\citenamefont {Carrasco}\ \emph {et~al.}(2021)\citenamefont
  {Carrasco}, \citenamefont {Rodina},\ and\ \citenamefont
  {Zekioglu}}]{Carrasco:2021ptp}%
  \BibitemOpen
  \bibfield  {author} {\bibinfo {author} {\bibfnamefont {John Joseph~M.}\
  \bibnamefont {Carrasco}}, \bibinfo {author} {\bibfnamefont {Laurentiu}\
  \bibnamefont {Rodina}}, \ and\ \bibinfo {author} {\bibfnamefont {Suna}\
  \bibnamefont {Zekioglu}},\ }\bibfield  {title} {\enquote {\bibinfo {title}
  {{Composing effective prediction at five points}},}\ }\href {\doibase
  10.1007/JHEP06(2021)169} {\bibfield  {journal} {\bibinfo  {journal} {JHEP}\
  }\textbf {\bibinfo {volume} {06}},\ \bibinfo {pages} {169} (\bibinfo {year}
  {2021})},\ \Eprint {http://arxiv.org/abs/2104.08370} {arXiv:2104.08370
  [hep-th]} \BibitemShut {NoStop}%
\bibitem [{\citenamefont {Pavao}(2023)}]{Pavao:2022kog}%
  \BibitemOpen
  \bibfield  {author} {\bibinfo {author} {\bibfnamefont {Nicolas~H.}\
  \bibnamefont {Pavao}},\ }\bibfield  {title} {\enquote {\bibinfo {title}
  {{Effective observables for electromagnetic duality from novel amplitude
  decomposition}},}\ }\href {\doibase 10.1103/PhysRevD.107.065020} {\bibfield
  {journal} {\bibinfo  {journal} {Phys. Rev. D}\ }\textbf {\bibinfo {volume}
  {107}},\ \bibinfo {pages} {065020} (\bibinfo {year} {2023})},\ \Eprint
  {http://arxiv.org/abs/2210.12800} {arXiv:2210.12800 [hep-th]} \BibitemShut
  {NoStop}%
\bibitem [{\citenamefont {Bern}\ \emph {et~al.}(2019)\citenamefont {Bern},
  \citenamefont {Carrasco}, \citenamefont {Chiodaroli}, \citenamefont
  {Johansson},\ and\ \citenamefont {Roiban}}]{BCJReview}%
  \BibitemOpen
  \bibfield  {author} {\bibinfo {author} {\bibfnamefont {Zvi}\ \bibnamefont
  {Bern}}, \bibinfo {author} {\bibfnamefont {John~Joseph}\ \bibnamefont
  {Carrasco}}, \bibinfo {author} {\bibfnamefont {Marco}\ \bibnamefont
  {Chiodaroli}}, \bibinfo {author} {\bibfnamefont {Henrik}\ \bibnamefont
  {Johansson}}, \ and\ \bibinfo {author} {\bibfnamefont {Radu}\ \bibnamefont
  {Roiban}},\ }\bibfield  {title} {\enquote {\bibinfo {title} {{The Duality
  Between Color and Kinematics and its Applications}},}\ }\href@noop {} {\
  (\bibinfo {year} {2019})},\ \Eprint {http://arxiv.org/abs/1909.01358}
  {arXiv:1909.01358 [hep-th]} \BibitemShut {NoStop}%
\bibitem [{\citenamefont {Bern}\ \emph
  {et~al.}(2022{\natexlab{b}})\citenamefont {Bern}, \citenamefont {Carrasco},
  \citenamefont {Chiodaroli}, \citenamefont {Johansson},\ and\ \citenamefont
  {Roiban}}]{Bern:2022wqg}%
  \BibitemOpen
  \bibfield  {author} {\bibinfo {author} {\bibfnamefont {Zvi}\ \bibnamefont
  {Bern}}, \bibinfo {author} {\bibfnamefont {John~Joseph}\ \bibnamefont
  {Carrasco}}, \bibinfo {author} {\bibfnamefont {Marco}\ \bibnamefont
  {Chiodaroli}}, \bibinfo {author} {\bibfnamefont {Henrik}\ \bibnamefont
  {Johansson}}, \ and\ \bibinfo {author} {\bibfnamefont {Radu}\ \bibnamefont
  {Roiban}},\ }\bibfield  {title} {\enquote {\bibinfo {title} {{The SAGEX
  Review on Scattering Amplitudes, Chapter 2: An Invitation to Color-Kinematics
  Duality and the Double Copy}},}\ }\href@noop {} {\  (\bibinfo {year}
  {2022}{\natexlab{b}})},\ \Eprint {http://arxiv.org/abs/2203.13013}
  {arXiv:2203.13013 [hep-th]} \BibitemShut {NoStop}%
\bibitem [{\citenamefont {Adamo}\ \emph {et~al.}(2022)\citenamefont {Adamo},
  \citenamefont {Carrasco}, \citenamefont {Carrillo-Gonz\'alez}, \citenamefont
  {Chiodaroli}, \citenamefont {Elvang}, \citenamefont {Johansson},
  \citenamefont {O'Connell}, \citenamefont {Roiban},\ and\ \citenamefont
  {Schlotterer}}]{Adamo:2022dcm}%
  \BibitemOpen
  \bibfield  {author} {\bibinfo {author} {\bibfnamefont {Tim}\ \bibnamefont
  {Adamo}}, \bibinfo {author} {\bibfnamefont {John Joseph~M.}\ \bibnamefont
  {Carrasco}}, \bibinfo {author} {\bibfnamefont {Mariana}\ \bibnamefont
  {Carrillo-Gonz\'alez}}, \bibinfo {author} {\bibfnamefont {Marco}\
  \bibnamefont {Chiodaroli}}, \bibinfo {author} {\bibfnamefont {Henriette}\
  \bibnamefont {Elvang}}, \bibinfo {author} {\bibfnamefont {Henrik}\
  \bibnamefont {Johansson}}, \bibinfo {author} {\bibfnamefont {Donal}\
  \bibnamefont {O'Connell}}, \bibinfo {author} {\bibfnamefont {Radu}\
  \bibnamefont {Roiban}}, \ and\ \bibinfo {author} {\bibfnamefont {Oliver}\
  \bibnamefont {Schlotterer}},\ }\bibfield  {title} {\enquote {\bibinfo {title}
  {{Snowmass White Paper: the Double Copy and its Applications}},}\ }in\
  \href@noop {} {\emph {\bibinfo {booktitle} {{2022 Snowmass Summer Study}}}}\
  (\bibinfo {year} {2022})\ \Eprint {http://arxiv.org/abs/2204.06547}
  {arXiv:2204.06547 [hep-th]} \BibitemShut {NoStop}%
\bibitem [{\citenamefont {Bern}\ \emph
  {et~al.}(2010{\natexlab{b}})\citenamefont {Bern}, \citenamefont {Carrasco},\
  and\ \citenamefont {Johansson}}]{Bern:2010ue}%
  \BibitemOpen
  \bibfield  {author} {\bibinfo {author} {\bibfnamefont {Zvi}\ \bibnamefont
  {Bern}}, \bibinfo {author} {\bibfnamefont {John Joseph~M.}\ \bibnamefont
  {Carrasco}}, \ and\ \bibinfo {author} {\bibfnamefont {Henrik}\ \bibnamefont
  {Johansson}},\ }\bibfield  {title} {\enquote {\bibinfo {title} {{Perturbative
  Quantum Gravity as a Double Copy of Gauge Theory}},}\ }\href {\doibase
  10.1103/PhysRevLett.105.061602} {\bibfield  {journal} {\bibinfo  {journal}
  {Phys. Rev. Lett.}\ }\textbf {\bibinfo {volume} {105}},\ \bibinfo {pages}
  {061602} (\bibinfo {year} {2010}{\natexlab{b}})},\ \Eprint
  {http://arxiv.org/abs/1004.0476} {arXiv:1004.0476 [hep-th]} \BibitemShut
  {NoStop}%
\bibitem [{\citenamefont {Cachazo}\ \emph {et~al.}(2015)\citenamefont
  {Cachazo}, \citenamefont {He},\ and\ \citenamefont {Yuan}}]{Cachazo2014xea}%
  \BibitemOpen
  \bibfield  {author} {\bibinfo {author} {\bibfnamefont {Freddy}\ \bibnamefont
  {Cachazo}}, \bibinfo {author} {\bibfnamefont {Song}\ \bibnamefont {He}}, \
  and\ \bibinfo {author} {\bibfnamefont {Ellis~Ye}\ \bibnamefont {Yuan}},\
  }\bibfield  {title} {\enquote {\bibinfo {title} {{Scattering equations and
  matrices: From Einstein to Yang-Mills, DBI and NLSM}},}\ }\href {\doibase
  10.1007/JHEP07(2015)149} {\bibfield  {journal} {\bibinfo  {journal} {JHEP}\
  }\textbf {\bibinfo {volume} {07}},\ \bibinfo {pages} {149} (\bibinfo {year}
  {2015})},\ \Eprint {http://arxiv.org/abs/1412.3479} {arXiv:1412.3479
  [hep-th]} \BibitemShut {NoStop}%
\bibitem [{\citenamefont {Elvang}\ \emph {et~al.}(2021)\citenamefont {Elvang},
  \citenamefont {Hadjiantonis}, \citenamefont {Jones},\ and\ \citenamefont
  {Paranjape}}]{Elvang:2020kuj}%
  \BibitemOpen
  \bibfield  {author} {\bibinfo {author} {\bibfnamefont {Henriette}\
  \bibnamefont {Elvang}}, \bibinfo {author} {\bibfnamefont {Marios}\
  \bibnamefont {Hadjiantonis}}, \bibinfo {author} {\bibfnamefont {Callum
  R.~T.}\ \bibnamefont {Jones}}, \ and\ \bibinfo {author} {\bibfnamefont
  {Shruti}\ \bibnamefont {Paranjape}},\ }\bibfield  {title} {\enquote {\bibinfo
  {title} {{Electromagnetic Duality and D3-Brane Scattering Amplitudes Beyond
  Leading Order}},}\ }\href {\doibase 10.1007/JHEP04(2021)173} {\bibfield
  {journal} {\bibinfo  {journal} {JHEP}\ }\textbf {\bibinfo {volume} {04}},\
  \bibinfo {pages} {173} (\bibinfo {year} {2021})},\ \Eprint
  {http://arxiv.org/abs/2006.08928} {arXiv:2006.08928 [hep-th]} \BibitemShut
  {NoStop}%
\bibitem [{\citenamefont {Carrasco}\ and\ \citenamefont
  {Rodina}(2019)}]{Carrasco:2019qwr}%
  \BibitemOpen
  \bibfield  {author} {\bibinfo {author} {\bibfnamefont {John Joseph~M.}\
  \bibnamefont {Carrasco}}\ and\ \bibinfo {author} {\bibfnamefont {Laurentiu}\
  \bibnamefont {Rodina}},\ }\bibfield  {title} {\enquote {\bibinfo {title} {{UV
  considerations on scattering amplitudes in a web of theories}},}\ }\href@noop
  {} {\  (\bibinfo {year} {2019})},\ \Eprint {http://arxiv.org/abs/1908.08033}
  {arXiv:1908.08033 [hep-th]} \BibitemShut {NoStop}%
\bibitem [{\citenamefont {Broedel}\ and\ \citenamefont
  {Dixon}(2012)}]{Broedel2012rc}%
  \BibitemOpen
  \bibfield  {author} {\bibinfo {author} {\bibfnamefont {Johannes}\
  \bibnamefont {Broedel}}\ and\ \bibinfo {author} {\bibfnamefont {Lance~J.}\
  \bibnamefont {Dixon}},\ }\bibfield  {title} {\enquote {\bibinfo {title}
  {{Color-kinematics duality and double-copy construction for amplitudes from
  higher-dimension operators}},}\ }\href {\doibase 10.1007/JHEP10(2012)091}
  {\bibfield  {journal} {\bibinfo  {journal} {JHEP}\ }\textbf {\bibinfo
  {volume} {10}},\ \bibinfo {pages} {091} (\bibinfo {year} {2012})},\ \Eprint
  {http://arxiv.org/abs/1208.0876} {arXiv:1208.0876 [hep-th]} \BibitemShut
  {NoStop}%
\bibitem [{\citenamefont {Carrasco}\ \emph {et~al.}(2013)\citenamefont
  {Carrasco}, \citenamefont {Kallosh}, \citenamefont {Roiban},\ and\
  \citenamefont {Tseytlin}}]{Carrasco:2013ypa}%
  \BibitemOpen
  \bibfield  {author} {\bibinfo {author} {\bibfnamefont {J.~J.~M.}\
  \bibnamefont {Carrasco}}, \bibinfo {author} {\bibfnamefont {R.}~\bibnamefont
  {Kallosh}}, \bibinfo {author} {\bibfnamefont {R.}~\bibnamefont {Roiban}}, \
  and\ \bibinfo {author} {\bibfnamefont {A.~A.}\ \bibnamefont {Tseytlin}},\
  }\bibfield  {title} {\enquote {\bibinfo {title} {{On the U(1) duality anomaly
  and the S-matrix of N=4 supergravity}},}\ }\href {\doibase
  10.1007/JHEP07(2013)029} {\bibfield  {journal} {\bibinfo  {journal} {JHEP}\
  }\textbf {\bibinfo {volume} {07}},\ \bibinfo {pages} {029} (\bibinfo {year}
  {2013})},\ \Eprint {http://arxiv.org/abs/1303.6219} {arXiv:1303.6219
  [hep-th]} \BibitemShut {NoStop}%
\bibitem [{\citenamefont {Bern}\ \emph {et~al.}(2018)\citenamefont {Bern},
  \citenamefont {Parra-Martinez},\ and\ \citenamefont {Roiban}}]{Bern:2017rjw}%
  \BibitemOpen
  \bibfield  {author} {\bibinfo {author} {\bibfnamefont {Zvi}\ \bibnamefont
  {Bern}}, \bibinfo {author} {\bibfnamefont {Julio}\ \bibnamefont
  {Parra-Martinez}}, \ and\ \bibinfo {author} {\bibfnamefont {Radu}\
  \bibnamefont {Roiban}},\ }\bibfield  {title} {\enquote {\bibinfo {title}
  {{Canceling the U(1) Anomaly in the $S$ Matrix of $N$=4 Supergravity}},}\
  }\href {\doibase 10.1103/PhysRevLett.121.101604} {\bibfield  {journal}
  {\bibinfo  {journal} {Phys. Rev. Lett.}\ }\textbf {\bibinfo {volume} {121}},\
  \bibinfo {pages} {101604} (\bibinfo {year} {2018})},\ \Eprint
  {http://arxiv.org/abs/1712.03928} {arXiv:1712.03928 [hep-th]} \BibitemShut
  {NoStop}%
\bibitem [{\citenamefont {Carrasco}\ \emph
  {et~al.}(2023{\natexlab{a}})\citenamefont {Carrasco}, \citenamefont
  {Lewandowski},\ and\ \citenamefont {Pavao}}]{Carrasco:2022lbm}%
  \BibitemOpen
  \bibfield  {author} {\bibinfo {author} {\bibfnamefont {John Joseph~M.}\
  \bibnamefont {Carrasco}}, \bibinfo {author} {\bibfnamefont {Matthew}\
  \bibnamefont {Lewandowski}}, \ and\ \bibinfo {author} {\bibfnamefont
  {Nicolas~H.}\ \bibnamefont {Pavao}},\ }\bibfield  {title} {\enquote {\bibinfo
  {title} {{Color-Dual Fates of F3, R3, and N=4 Supergravity}},}\ }\href
  {\doibase 10.1103/PhysRevLett.131.051601} {\bibfield  {journal} {\bibinfo
  {journal} {Phys. Rev. Lett.}\ }\textbf {\bibinfo {volume} {131}},\ \bibinfo
  {pages} {051601} (\bibinfo {year} {2023}{\natexlab{a}})},\ \Eprint
  {http://arxiv.org/abs/2203.03592} {arXiv:2203.03592 [hep-th]} \BibitemShut
  {NoStop}%
\bibitem [{\citenamefont {Carrasco}\ \emph
  {et~al.}(2023{\natexlab{b}})\citenamefont {Carrasco}, \citenamefont
  {Lewandowski},\ and\ \citenamefont {Pavao}}]{Carrasco:2022sck}%
  \BibitemOpen
  \bibfield  {author} {\bibinfo {author} {\bibfnamefont {John Joseph~M.}\
  \bibnamefont {Carrasco}}, \bibinfo {author} {\bibfnamefont {Matthew}\
  \bibnamefont {Lewandowski}}, \ and\ \bibinfo {author} {\bibfnamefont
  {Nicolas~H.}\ \bibnamefont {Pavao}},\ }\bibfield  {title} {\enquote {\bibinfo
  {title} {{Double-copy towards supergravity inflation with
  \ensuremath{\alpha}-attractor models}},}\ }\href {\doibase
  10.1007/JHEP02(2023)015} {\bibfield  {journal} {\bibinfo  {journal} {JHEP}\
  }\textbf {\bibinfo {volume} {02}},\ \bibinfo {pages} {015} (\bibinfo {year}
  {2023}{\natexlab{b}})},\ \Eprint {http://arxiv.org/abs/2211.04441}
  {arXiv:2211.04441 [hep-th]} \BibitemShut {NoStop}%
\end{thebibliography}%

\end{document}